\pdfoutput=1
\documentclass{crasty}
\usepackage{cite}
\usepackage{float}

\def\gsim{\mathrel{\rlap{\lower4pt\hbox{\hskip1pt$\sim$}}}}

\usepackage[]{lineno}

\begin{document}
%
%
\noindent
LHeC-Note-2012-005 GEN ~~~~~~~~~~\\
Geneva, October 15, 2012
%
\begin{figure}[h]
\vspace{-2.cm}
\hspace{13.5cm}
\includegraphics[clip=,width=.15\textwidth]{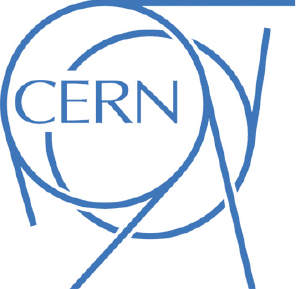}
\end{figure}
\begin{figure}[h]
\vspace{-1.3cm}
\hspace{4.3cm}
\includegraphics[clip=,width=0.45\textwidth]{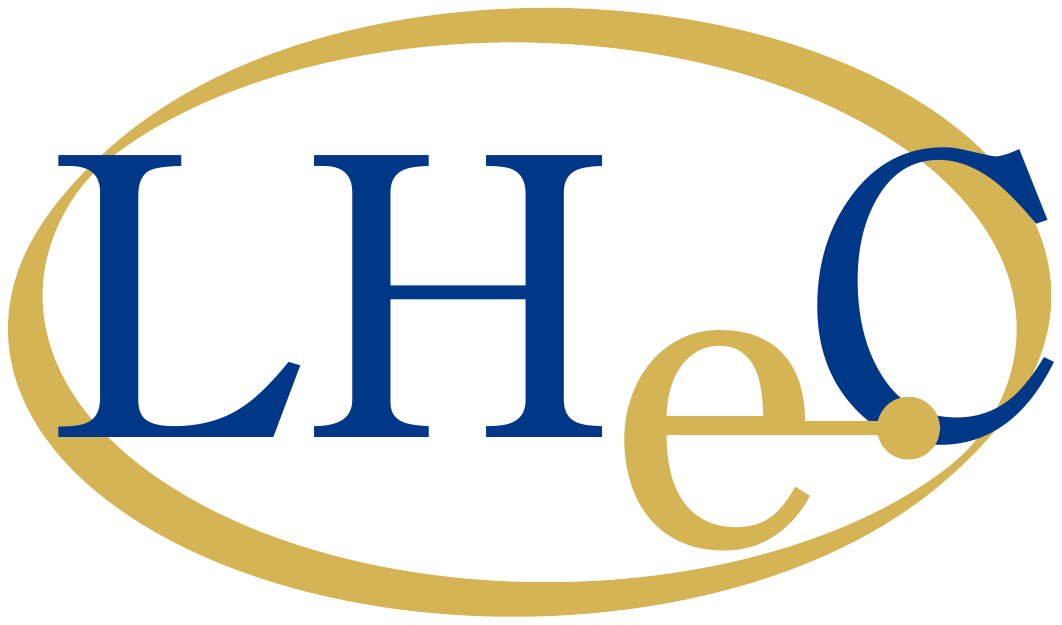}
\end{figure}
\begin{center}
\vspace{-.5cm}
\begin{LARGE}
\bf{On the Relation of the LHeC and the LHC} \\
\end{LARGE}
\vspace{0.5cm}
\begin{LARGE}
 \end{LARGE}
 \vspace{1cm}
%
\begin{Large}
\bf{LHeC Study Group
\footnote{For the current author list
see the end of this note. Contacts: 
oliver.bruning@cern.ch and max.klein@cern.ch}
}\\ 
\end{Large}
\vspace{0.5cm}
$\bf{Abstract}$ \\
\vspace{0.3cm}
\end{center}
The present note relies on the recently published 
conceptual design report of the LHeC and extends
the first contribution to the European strategy
debate in emphasising the role of the LHeC to complement
and complete the high luminosity LHC programme.
The brief discussion therefore focuses on the importance
of high precision PDF and $\alpha_s$ determinations
for the physics beyond the Standard Model (GUTs,
SUSY, Higgs).  
Emphasis is also given to the importance  
of high parton density phenomena in nuclei and their relevance
to the heavy ion physics programme at the LHC.
\\
\vspace{1cm}
\begin{center}
Submitted to the European Strategy for Particle Physics, \\
October 2012.
\end{center}
%
\newpage
%
%
\section{Introduction}
\label{intro}
Deep inelastic lepton-hadron scattering is the cleanest and most
precise probe of parton dynamics in protons and nuclei.
The LHeC is the only current proposal for TeV-scale lepton-hadron 
scattering and the only medium-term potential complement to the
LHC $pp$, $AA$ and $pA$ programme at the energy frontier. As such,
it has a rich and diverse physics programme of its own, as documented
extensively in the recent conceptual design report (CDR)
\cite{AbelleiraFernandez:2012cc}
and summarised in an initial submission by the LHeC Study Group
to the European Strategy of Particle Physics (ESPP) discussion
prior to the Cracow Symposium \cite{cracow1}.
The focus of this second submission to the ESPP is a 
further exploration of the relationship between the LHeC and the LHC. 
Specifically, by improving the understanding of the LHC initial state through 
tighter constraints on parton densities and providing information
on many other aspects of strong interactions, the LHeC 
extends the capabilities of the LHC programme substantially. 

%

The LHeC offers the
prospect of synchronous operation of a new $ep$ (and $eA$)
facility with 
the high luminosity phase of
the LHC (HL-LHC). The physics programme 
of the GPD experiments on this timescale, as illustrated in the ATLAS and
CMS contributions to the ESPP process, is centred around
two major objectives: 
i) the most precise and complete
exploration possible of the newly discovered Higgs-like particle and 
ii) maximising the sensitivity of the continuing search for
new particles and symmetries or extra dimensions in the
few TeV range of mass.
This document investigates the potential impact
of precision LHeC results on these objectives, as we understand them 
at present, recognising that the situation will 
evolve with time and 
deserves continuous further study. 
It also recalls the
importance of deep inelastic electron-ion scattering
for the completion of the LHC heavy ion programme.
The mutual relations between the LHeC and the LHC
are of course much deeper than can be covered in 
a brief communication such as this.

As documented in detail in \cite{AbelleiraFernandez:2012cc},
the parton density (PDF) determinations offered by the LHeC are substantially
superior to the possibilities using LHC data alone and, for the first
time, provide a full flavour decomposition essentially
free of assumptions. The LHeC also promises a broad and
unique programme of further strong interaction physics,
such as the exploration of a newly accessed
high density, low coupling regime at low $x$
and a new level of precision and hugely extended 
kinematic coverage on the partonic structure of nuclei.
Combined with competitive  sensitivity to new
physics in channels where initial state lepton quantum numbers are an advantage,
the LHeC represents a cost effective means of fully exploiting the LHC and
substantially extending its physics programme.

Following the discovery by 
ATLAS~\cite{:2012gk} and CMS~\cite{:2012gu} 
of a new boson
with a mass of about $126$\,GeV, 
there are two main aims for the future of the field.
Firstly, to determine whether this particle is the 
Standard Model Higgs boson or something more exotic and, 
secondly, to extend as far as possible the range of sensitivity
for the discovery of other new particles. 
ATLAS and CMS have reported exclusion limits for a wide range
of massive new particles in the $1 - 2 \ {\rm TeV}$ range. 
With no strong evidence for new effects so far,
the need to further extend such searches to the largest masses possible
is paramount. The increased beam energy
following LS1 will provide a first major step in this direction. 
Beyond that, further progress is limited by luminosity, due to the 
fast-falling cross sections as higher and higher $x$ PDFs are
involved (especially for gluon initiated processes), and by the 
uncertainties on those PDFs. To fully exploit the new particle
discovery range of the LHC,
both a luminosity upgrade and tighter external 
PDF constraints are therefore required.

The future exploration of the Higgs sector at the LHC,
for example by measuring
relative couplings and testing the CP structure, may similarly
become limited by theory uncertainties derived from PDF measurements
once the very high luminosities possible at HL-LHC have been 
accumulated \cite{ATLAS-LHLHC}. This is particularly true for the 
workhorse channels in which the Higgs decays to $\gamma \gamma$
or four charged leptons. Whilst LHC inclusive $W$ and $Z$ production
data will somewhat improve constraints on the quark densities of the proton
at the electroweak scale, they
will have a limited impact on the gluon density, which is more pertinent
to Higgs physics, given the dominant $gg$ production mechanism. 

The present document is organised as follows. Section~2 presents a 
brief reminder of the Linac-Ring configuration of the LHeC,
which is being prepared via prototype and design
developments with a view to endorsement of the project in the
next few years. This includes a brief exploration of possible ways of
enhancing the luminosity to the level of $10^{34}$\,cm$^{-2}$s$^{-1}$.
Section~3 is devoted to PDF determinations, including the relation between
$pp$ and $ep$, the LHeC's potential to
unfold all quark flavours and to unfold the gluon
density to an unprecedented level of precision in a huge $x$ range. 
Section~4 provides a discussion of two key
features of the LHeC which may be 
crucial in the establishment of new theories or
particles: firstly the prospect of measuring $\alpha_s$
to per mille accuracy and secondly the precision calculation of
SUSY cross sections involving high $x$ partons at the limits of the 
accessible HL-LHC
mass range, 
gluino pair production being used as an example.
Section~5 discusses
the main contributions of the LHeC to Higgs physics;
its potential to find new physics at the cleanly accessible
$WWH$ (and $ZZH$) vertices and the corresponding
precision coupling measurements to vector bosons, 
as well as its reduction of the QCD-dominated
theory uncertainties on Higgs production in $pp$ collisions. 
Finally, in Section~6, the
importance of $eA$ DIS measurements for the LHC heavy ion programme is 
briefly reviewed.

%
%
\section{The LHeC Linac-Ring Collider}
\label{linac}
%


The LHeC aims at colliding the high-energy protons and heavy ions 
circulating in the LHC with 60-GeV polarised electrons 
and possibly also positrons. 
The LHeC is realised by adding to the LHC a separate 9-km 
racetrack-shaped recirculating superconducting (SC) 
energy-recovery linac (ERL).
The key components of the LHeC are the two 1-km 10-GeV SC linacs of the ERL,
comparable in scale to the 17.5-GeV SC linac 
of the European XFEL presently under construction. 
The LHeC ERL provides a design  
lepton beam current of 6.4 mA at the $ep$ collision point,  
which is taken to be at IP2 of the LHC.   
Aside from the IP2 interaction region, the LHeC underground infrastructure 
is fully decoupled from the existing LHC tunnel. 
Two of the access shafts could be located on the CERN Prevessin site. 
 
The LHeC is designed to operate with simultaneous LHC 
$pp$ (or $AA$) collisions.
LHeC operation is fully transparent to the other LHC experiments thanks to 
the low lepton bunch charge and resulting minuscule beam-beam tune shift 
experienced by the protons,  
together with the choice of the LHeC circumference 
to be equal to a third of the LHC's in order  
to allow for ion-clearing gaps in the ERL without perturbing LHC steady-state operation.  


\begin{table*}[htb]
\begin{center}
\begin{tabular}{|l|cc|}
\hline
parameter [unit] &  \multicolumn{2}{|c|}{LHeC} \\ 
\hline
species & $e$ & $p$, $^{208}  $Pb$^{82+}$ \\    
beam energy (/nucleon) [GeV] & 60 & 7000, 2760 \\
bunch spacing [ns] & 25, 100 & 25, 100 \\
bunch intensity (nucleon) [$10^{10}$]  & 0.1 (0.2), 0.4 & 17 (22), 2.5  \\
beam current [mA] & 6.4 (12.8) & 860 (1110), 6 \\
rms bunch length [mm] &  0.6 & 75.5  \\
polarisation [\%] & 90 ($e^{+}$ none) & none, none \\  
normalised rms emittance [$\mu$m]  & 50  & 3.75 (2.0), 1.5  \\
geometric rms emittance [nm] & 0.43  & 0.50 (0.31)  \\
IP beta function $\beta_{x,y}^{\ast}$ [m] & 0.12 (0.032) & 0.1 (0.05) \\   
IP spot size [$\mu$m]  & 7.2 (3.7) & 7.2 (3.7)\\   
synchrotron tune $Q_{s}$ &  --- & $1.9\times 10^{-3}$ \\ 
hadron beam-beam parameter &   \multicolumn{2}{|c|}{0.0001 (0.0002)} \\ 
lepton disruption parameter $D$  & \multicolumn{2}{|c|}{6 (30)} \\ 
crossing angle &   \multicolumn{2}{|c|}{0 (detector-integrated dipole)} \\ 
hourglass reduction factor $H_{hg}$  & \multicolumn{2}{|c|}{0.91 (0.67)} \\ 
pinch enhancement factor $H_{D}$  & \multicolumn{2}{|c|}{1.35 (0.3 for $e^{+}$)} \\ 
CM energy [TeV] &  \multicolumn{2}{|c|}{1.3, 0.81} \\ 
luminosity / nucleon [$10^{33}$ cm$^{-2}$s$^{-1}$] &  \multicolumn{2}{|c|}{1 (10), 0.2} \\ 
\hline
\end{tabular}
\end{center}
\caption{LHeC parameters. The numbers give
 the default values with optimum values for maximum
 $ep$ luminosity in parenthesis and values
 for the $ePb$ configuration separated by a comma.}
\label{tab1}
\end{table*}

LHeC has been designed under the constraint that 
the total electrical power for the LHeC lepton branch 
should not exceed $100$\,MW 
(about half the present maximum CERN site power).  
The LHeC electrical power budget is dominated by the RF and by the 
cryo power for the two 1-km long SC linacs. 
The cryo power required and, therefore, also the  
size of the cryoplants (as well as the maximum lepton current) 
are directly linked to the unloaded quality factor of the cavities, $Q_0$. 
With a $Q_{0}$ of $2.5\times 10^{10}$, the total  
main-linac cryopower amounts to 23 MW. The RF power needed for 
RF microphonics control is about 24 MW, and the 
extra-RF power needed for compensating SR losses 
at 6.4-mA current also to 23 MW. 
The remaining components, like injectors 
or arc magnets, require a few MW each.  

Together with rather conservative assumptions for most parameters, 
the $100$\,MW power limit yields  
the LHeC $ep$ target luminosity of $10^{33}$~cm$^{-2}$s$^{-1}$. 
However, extensions to significantly higher luminosity, e.g., 
$10^{34}$~cm$^{-2}$s$^{-1}$, are possible by a 
combination of improvements, namely   
(1) by considering normalised proton beam emittances of 2 $\mu$m
(as achieved in 2011/12 LHC operation) instead of
3.75~$\mu$m; (2) by a further reduction of the 
proton IP beta function from 0.1\,m down to 0.05\,m,
which should be possible by using a variant of 
the so called ATS optics;  
(3) by increasing the proton bunch intensity  
from $1.7\times 10^{11}$ to the HL-LHC
25\,ns target value of $2.2\times 10^{11}$ 
[for the 50-ns HL-LHC scenario it would be even 
$3.3\times 10^{11}$ with a possible further factor 2.5 
increase of luminosity, to more than $2\times 10^{34}$~cm$^{-2}$s$^{-1}$]; 
and (4) by doubling the lepton beam current, 
which should be possible without exceeding 
the 100-MW power limit if the unloaded 
$Q_{0}$ value of the SC RF cavities can be raised to 
$4\times 10^{10}$ (as it is assumed for the similar eRHIC design). 
Table \ref{tab1} shows LHeC parameters, including, in parentheses,  
values for a higher-luminosity variant. 

The LHeC represents an interesting possibility for 
further efficient exploitation of the LHC infrastructure investment.
The development of a CW SC recirculating 
energy-recovery linac for LHeC would 
prepare for many possible future projects, e.g., 
for an International Linear Collider, 
for a neutrino factory, for a proton-driven plasma wake field accelerator,
or for a muon collider. 
With some additional arcs, using 4 instead of 3 passes 
through the linacs, a machine like the LHeC ERL (without energy recovery) 
could also operate as a Higgs factory $\gamma\gamma$ collider (SAPPHiRE). 

%
%
\section{Parton Distributions}
\subsection{PDFs in pp and ep}
\label{pdfs}
\label{pdftheory}
The factorisation theorems of perturbative QCD (see
Ref.~\cite{Collins:2011zzd} and references therein) express physical
observables for hard processes characterised by a large momentum scale
as the convolution of a perturbatively computable
partonic cross-section, determined 
in terms of the degrees of freedom of the QCD Lagrangian
--- quarks and gluons --- and universal, process--independent parton
distributions, up to corrections suppressed by powers of the
ratio of a characteristic QCD scale $\Lambda\sim100$~MeV to the large scale.

Factorisation is most rigorously established for deep inelastic
lepton-hadron scattering, where the expression for the
cross-section has schematically the form
\begin{equation}\label{eq:disfact}
\sigma_{DIS}(x,Q^2)=\int_x^1\frac{dz}{z}\,\hat\sigma_{DIS}(z,\alpha_s(Q^2))
f\left(Q^2,\frac{x}{z}\right),
\end{equation}
where $\sigma_{DIS}(x,Q^2)$ denotes generically a deep inelastic
 structure function, $Q^2$ is the hard scale of the process, namely,
 the virtuality of the gauge boson which mediates the scattering
 process, and the scaling variable is
 $x=Q^2/2p\cdot q$ in terms of the incoming hadron momentum $p$ and
 momentum transfer $q=k^\prime-k$ between the incoming ($k$) and
 outgoing ($k^\prime$) lepton momenta. In Eq.~(\ref{eq:disfact}), the
 partonic cross section $\hat\sigma_{DIS}(z,\alpha_s(Q^2))$ can be
 computed as a perturbative expansion in $\alpha_s(Q^2)$, while
 $f\left(Q^2,x\right)$ is a parton distribution; a sum over different
 kinds of partons (individual quark flavours and gluons) is
 understood but omitted for simplicity.
Equation~(\ref{eq:disfact}) directly follows from the Operator-Product
 Expansion, though it can also be derived from the computation of
 parton Feynman diagrams. 

Factorisation for hadronic cross-sections, such as for example the
production of an electroweak final state such as a Higgs or a $W$ or
$Z$ takes the form
\begin{equation}\label{eq:hadfact}
\sigma_{DY}(\tau,M^2)=\int_\tau^1\frac{dz}{z}\,\hat\sigma_{DY}(z,\alpha_s(M^2))
{\cal L}\left(\frac{\tau}{z}\right),
\end{equation}
where now $\sigma_{DY}(\tau,M^2)$ and
$\hat\sigma_{DY}(z,\alpha_s(M^2))$ are respectively hadronic and
partonic 
cross sections
for production of the electroweak final state of mass $M$: with the hadronic
$\sigma$ measurable, and the partonic $\hat \sigma$ computable in
perturbation theory, and  the scaling variable is
$\tau=\frac{M^2}{s}$.
The dependence on parton distributions now goes through the parton luminosity
\begin{equation}\label{eq:lumi}
{\cal L}(\tau)=\sum_{a,b} \int_{\tau}^1 \frac{dx}{x}  f_{a/h_1}(x) 
f_{b/h_2}(\tau/x),
\end{equation}
which depends on the parton distribution from parton $a$ and $b$,
respectively, extracted from the two incoming hadrons $h_1$ and $h_2$. Again, we
have neglected a sum over partonic channels for simplicity.

Because taking a Mellin transform turns the convolutions in
Eqs.(\ref{eq:disfact}-\ref{eq:hadfact}) into ordinary products, the
universality of PDFs implies that one can form a PDF-independent
measurable ratio:
\begin{equation}\label{eq:pdfindep}
\frac{\sigma_{DY}(N,M^2)}{\sigma^2_{DIS}(N,M^2)}=\frac{\hat\sigma_{DY}(N,M^2)}{\hat\sigma_{DIS}^2(N,M^2)},
\end{equation}
where  $\sigma_{DY}(N,M^2)=\int_0^1d\tau\,\tau^{N-1}\sigma(\tau,M^2)$,
and analogously for the other cross sections. It is the possibility of
constructing such a ratio, which only depends on perturbatively
computable partonic cross sections, which guarantees predictivity
of Drell-Yan processes for example.

Factorisation for hadronic processes in which a colourless electroweak
final state is produced, such as Higgs, or a real or virtual  $W$, $Z$
or $\gamma$ is firmly established on the basis of an all-order
analysis of the relevant Feynman diagrams. Furthermore, such processes
are currently available up to NNLO in perturbation theory at the
differential level and partly at N$^3$LO for total cross sections: no
counterexample to factorisation has been found up to this order.
Factorisation is also well-established for sufficiently inclusive
coloured final states, such as the one-jet and dijet cross section. In
this case no fully rigorous all-order argument is really available,
but no counterexample to factorisation has ever been found either.

Recent global determinations of parton distributions, such as CTEQ~\cite{Lai:2010vv},
MSTW~\cite{Martin:2009iq} or NNPDF~\cite{Ball:2011uy} combine both
deep inelastic scattering data with a variety of beams (electrons,
muons, neutrinos) and targets (protons, deuterons and heavier nuclei)
as well as hadron-hadron data such as virtual photon (Drell-Yan)
production, $W$ and $Z$ production, and single jet production. 
A typical such analysis includes over 2000 deep inelastic data points,
and over 1000 hadronic data points. It thereby provides both a test of
the factorisation framework which guarantees the mutual consistency of
these data, and a possibility of assessing their relative impact.

\begin{figure}[htb]
\begin{center}
\vspace{-5cm}
\includegraphics[clip=,width=.49\linewidth]{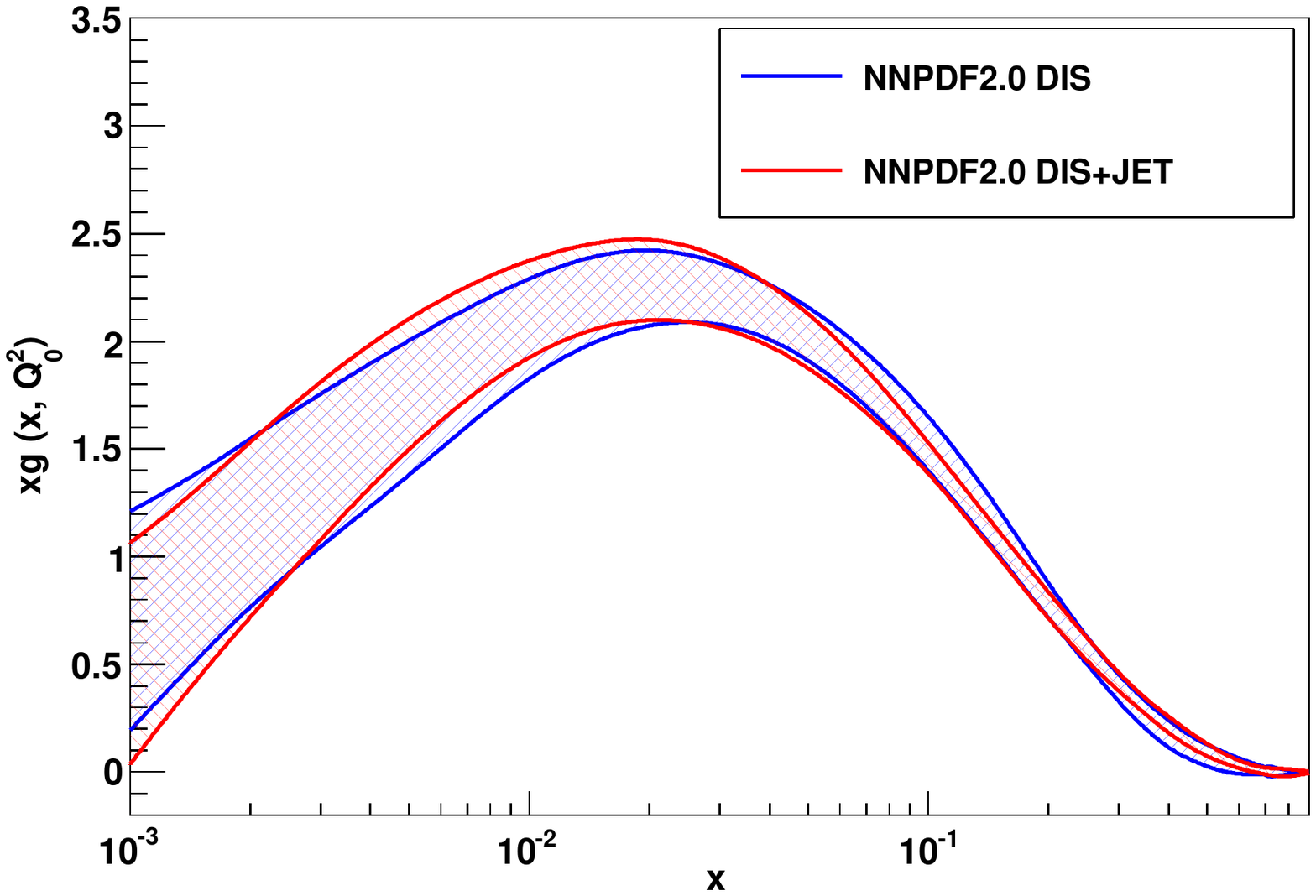}
\includegraphics[clip=,width=.49\linewidth]{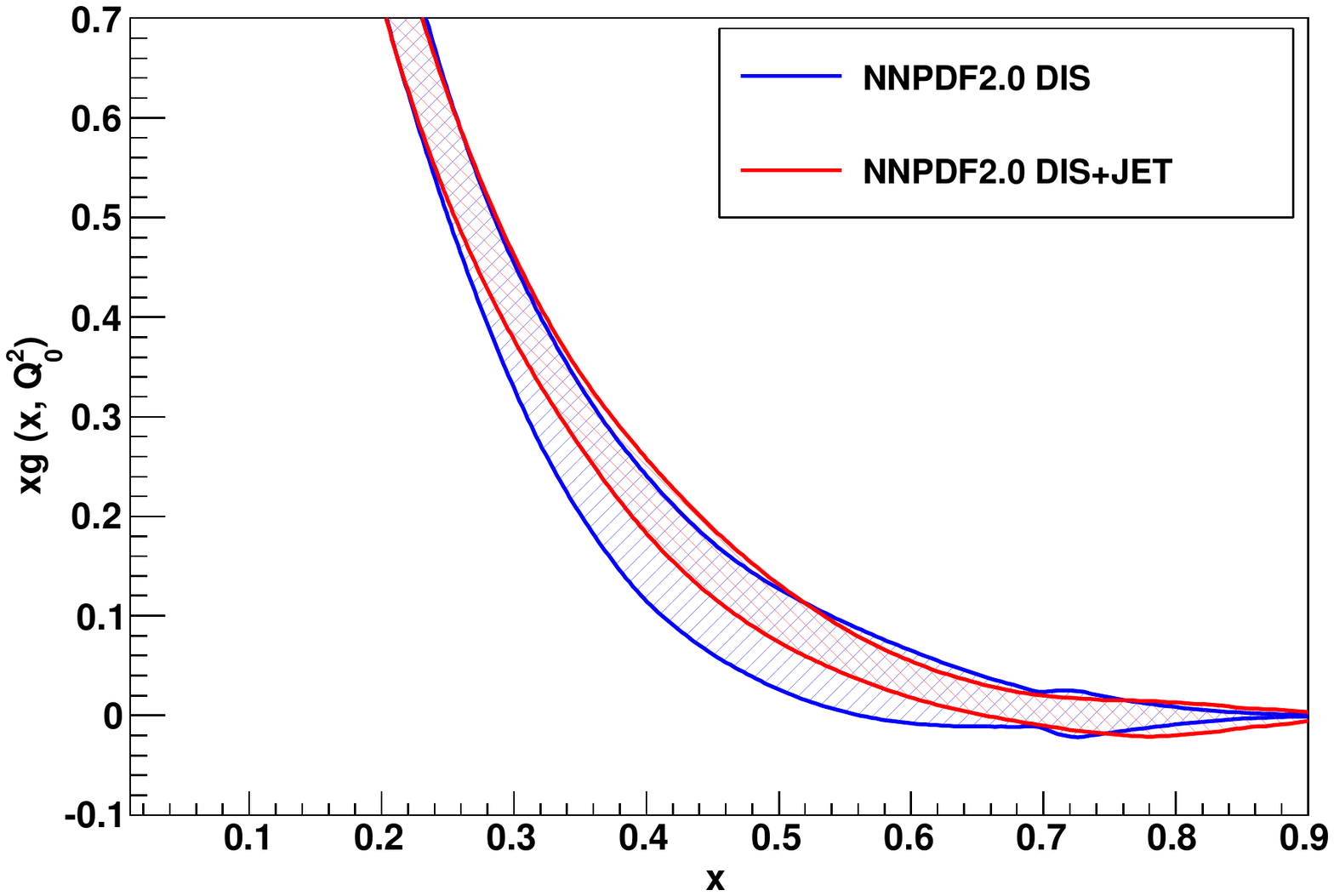}
\caption{The gluon distribution determined using deep inelastic
scattering data only (blue) or also including jet data (red), plotted
on a logarithmic (left) and linear (right) scale vs $x$ at
$Q^2=2$~GeV$^2$ (from~\cite{Forte:2010dt}).
}
\label{fig:jetdisconsist}
\end{center}
\end{figure}
As an example of both, Figure~\ref{fig:jetdisconsist} compares the
gluon distribution extracted by only fitting deep inelastic scattering
data, and
by also adding jet data to the DIS dataset. Such a comparison provides an
interesting test of factorisation for at least three reasons. First,
because one is comparing a lepton-hadron and a hadron-hadron
process. Second, because the gluon is determined through rather
different physical mechanisms: in deep inelastic scattering, where it
does not couple to the leading-order perturbative process because it
carries no electric or weak charge, the gluon is determined
by the $Q^2$ evolution. In jet production, instead, it
directly contributes both to the initial and final state. Finally,
because the jet data are taken at scales which are typically one or
two orders of magnitude larger than that of the deep inelastic data,
and thus the parton distribution extracted from one process has to be
evolved through renormalisation group equations in order to be  used
in the other process. Nevertheless, it is apparent from    
Fig.~\ref{fig:jetdisconsist}  that the gluon determined by
deep inelastic scattering and jets is consistent: when
adding jet to deep inelastic 
data, the central value moves very little, while the
uncertainty is somewhat reduced. It is also apparent from the
figure that the bulk of the information on the gluon distribution is
coming from deep inelastic scattering, with the jet data only
providing some useful but no determinant extra 
information.

Both these situations are generic, and likely to persist in the
future. On the one hand, perturbative factorisation is now firmly
established for a variety of processes, and electron-proton data can thus
reliably provide input for hadron collider processes. In fact, because
factorisation is most reliably established in deep inelastic
scattering, it is the availability of precise parton distributions
from lepton-hadron scattering which may allow detailed tests of the
validity of factorisation for  processes for which it is less 
well established. On the other hand, because lepton-hadron data are
generally subject
to smaller power-suppressed corrections, perturbatively more stable,
easier to compute than most hadronic processes so results to the
highest perturbative orders are available, and finally free of many
complications which arise when dealing with hadronic initial and final
states (such as jet definitions, or underlying event), 
lepton-proton data always provide a comparatively  more
competitive and theoretically reliable
determination of parton distributions than hadron-hadron data. The
natural scenario is one in which lepton-proton data are used to
determine parton distributions, and the latter are then used for
hadron collider processes, and there are strong reasons of principle
why this is the case.

\subsection{NC and CC Cross Section Measurements}
\label{crosec}
The determinations of parton distributions at the LHeC are of unique
range and quality because of a number of salient features which
characterise this experiment, especially with respect to HERA:
i) The LHeC greatly extends the kinematic range compared to HERA. The
increase in negative momentum transfer squared $Q^2$ is from
a maximum of about $0.03$ at HERA to $1$\,TeV$^2$ at the LHeC, and 
in $x$, e.g. for $Q^2 = 3$\,GeV$^2$,
from about $4 \cdot 10^{-5}$ to $2 \cdot 10^{-6}$.
ii) The projected increase of integrated luminosity
by a factor of $100$ allows to also extend the 
kinematic range at large $x$, from practically
about $0.4$ to $0.8$ in charged currents (CC).
This enables a precision mapping of the
high $x$ region, corresponding to large masses, of a few TeV,
in Drell-Yan scattering at the LHC.
iii) The increase in $Q^2$ implies that all parts of the neutral current
cross section, due to pure photon and pure $Z$ exchange, and their 
interference become of equal strength. This, combined with high
precision and CC data in a large kinematic range, enables a complete
separation of sea and valence quarks. It is crucial to understand that
such a basis of PDF determinations will render all previous
PDF determinations of inferior importance and practically reduce
any parameterisation uncertainty in QCD PDF fits to a negligible
level of importance. 

The superior nature of 
the DIS process for testing partons, with respect to Drell-Yan scattering,
the higher precision in $ep$ wrt $pp$ and the availability of
an enormous range in $Q^2$ for fixing parton evolution, as opposed
to the $Q^2 \simeq M_{W,Z}^2$ scale of the most accurate DY process at the LHC,
these and further features make the LHeC the appropriate
machine for transforming the LHC into a precision QCD, search
and Higgs factory in the twenties.

The analysis of PDF measurements of the LHeC has been based
on a full simulation of the NC and CC inclusive cross section measurements.
The assumptions on sources of systematic uncertainty
 are listed in Table\,\ref{tab:sys}~\cite{AbelleiraFernandez:2012cc}.
Broadly speaking, it is assumed
that with a new detector and high luminosity the H1 level of systematic
uncertainty is reached and improved by up to a factor of two.
\begin{table}[h]
  \centering
  \begin{tabular}{|l|c|}
    \hline
source of uncertainty & error on the source or cross section \\ \hline
scattered electron energy scale $\Delta E_e' /E_e'$ & 0.1 \% \\
scattered electron polar angle  & 0.1\,mrad \\
hadronic energy scale $\Delta E_h /E_h$ & 0.5\,\% \\
calorimeter noise (only $y < 0.01$) & 1-3\,\% \\ 
radiative corrections & 0.5\% \\
photoproduction background (only $y > 0.5$) & 1\,\% \\
global efficiency error & 0.7\,\%  \\
 \hline
  \end{tabular}
\caption{
Assumptions used in the simulation of the NC cross sections
on the size of uncertainties from various sources. 
These assumptions correspond to typical best values
achieved in the H1 experiment. The total cross section error
due to these uncertainties, e.g. for $Q^2 = 100$\,GeV$^2$,
is about $1.2$, $0.7$ and $2.0$\,\% for $y=0.84,~0.1,~0.004$.
}
\label{tab:sys}
\end{table}
The measurements of PDFs at the LHeC are complemented by high 
precision measurements of the heavy flavour quark densities
owing to a small beam spot, high luminosity and modern 
Silicon tracking techniques of high precision and wide acceptance.
An important part of the PDF programme at the LHeC is due
to the projected run with deuterons, which extends the knowledge
of neutron structure from DIS by $4$ orders of magnitude in
kinematic range. It needs deuterons to measure a singlet combination
of parton densities and to unfold the light sea flavour composition 
at low $x$. The LHeC is also an ideal and necessary configuration
to determine the nuclear PDFs for the first time in most of the 
kinematic range as is emphasised below. 

\subsection{Valence and Sea Quarks}
\label{heavyQ}
The LHeC is in a unique position to 
unravel all quark densities
in the proton with a complete quark flavour
separation for the first time and 
with unprecedented precision.
The huge 
phase space covered matches the needs of the LHC
and includes the extreme values of Bjorken $x$,
lowest, $x \simeq 10^{-5}$, where saturation
may set in and largest, near to $1$,
which determine the multi-TeV BSM
cross sections at the LHC.
The detailed shape measurements of the
various parton distributions, for example
of the strange density versus $Q^2$ and $x$,
imply that the currently large uncertainties
due to PDF parameterisations in pQCD fits
will be drastically reduced. The complete basis
for PDFs that the LHeC promises to deliver
is likely to lead to significant deviations in many places 
from the canonical PDF pattern we know today.
This is a necessary input for future LHC
measurements, such as  precision $Higgs$ coupling
and cross section determinations. There could also be 
discoveries made in deviations from the conventional PDF pattern, such as
the possible observation of anti-quarks 
to be different from their sea quarks
or an intrinsic heavy flavour component. Since 
the momentum is conserved, and shared
between quarks and gluons, any deviation
affects the overall pattern, which reflects
on other parts of physics~\footnote{
A recent example is the ATLAS observation of
the light sea to be flavour symmetric. Combined
with the precision HERA $F_2$ data this changes
the singlet sea by $8$\,\%, which has consequences
for the ultra high energy neutrino-nucleon
scattering cross sections.}. Moreover, 
a crucial variety of non-canonical PDFs
will be accessed: generalised, unintegrated,
diffractive, neutron, photon and nuclear 
parton distributions.

The basis for LHC physics and BSM discoveries
is QCD at high orders and 
the accurate knowledge of the 
classic PDFs. In the following,
based on~\cite{AbelleiraFernandez:2012cc},
some brief remarks are made on the unique
potential of the LHeC in the determination of 
the complete set of quark densities, while the mapping
of the gluon density is described subsequently below.
\begin{itemize}
\item
{\bf Valence quarks:}
The knowledge of the valence quark distributions, both
at large and at low Bjorken $x$, as derived in the current 
world data QCD fit analyses is amazingly limited.
An impressive improvement is expected from the LHeC. 
A NLO QCD fit to simulated inclusive neutral and charged
current LHeC data (see~\cite{AbelleiraFernandez:2012cc}) 
shows that the uncertainty of the down valence quark
distribution at, for example, $x=0.7$ can be reduced from
a level of   $50 - 100$\,\%  to about $5$\,\%.
This will be crucial for searches of new physics at the LHC
at the high energy frontier, in order to verify any excess
(or deficiency) compared to the SM prediction.
Direct access to valence quarks down to low $x\sim 0.001$ can be obtained at LHeC from
the  NC, $Z$ exchange related $e^{\pm}p$ cross section difference,
which can resolve possible sea-antiquark differences.
\item
{\bf Light sea quarks:}
The measurement of the structure functions $F_2 \propto 4U+D$, in $ep$
and $F_2 \propto U+D$, in $eD$ is
the basis for determining the light sea quark densities
in the nucleon.
LHeC will extend greatly the HERA kinematic coverage
to much lower $x$ and to higher 
scales $Q^2$.
%
From NC and CC measurements and comparing $ep$ with $eD$
data, the up and down sea quark densities will
be unfolded, which nowadays are assumed to be equal
at $x < 0.01$.
\item
{\bf Strange:} Several long-standing questions
are related to the strange quark density in the proton:
how much is it suppressed with respect to the other two light quarks?
Is there an asymmetry between the strange and anti-strange density?
The knowledge of the strange-quark density itself is important
for many processes, for instance for the  precision measurement
of the $W$ boson mass.
Information on the strange quark density is available from several experiments, 
in particular from previous Neutrino DIS experiments\footnote{
The interpretation of these neutrino data is sensitive to uncertainties
from charm quark fragmentation and nuclear corrections.},
but overall there is no real
understanding of the strange quark distribution.
The strange quark distribution is accessible at LHeC in charged current
scattering through the subprocesses $W^+s \to c$ (for positron beams)
and $W^-\bar{s} \to \bar c$ (for electron beams),
using charm tagging in the final state.
The LHeC simulation studies show that for the first time
precise measurements of the $s$ and $\bar{s}$ densities can be performed
over a large kinematical phase space in $x$ and $Q^2$.
\item
{\bf Charm:} 
Information on the charm content in the proton can 
be accessed at LHeC by measuring the inclusive
charm production cross section in neutral current DIS.
At low scales $Q^2 \sim m_c^2$ (with $m_c$ being the charm quark mass)
charm production has to be treated as being fully massive,
i.e. the charm quarks can only be dynamically produced 
in the reaction $\gamma g \rightarrow c\bar{c}$ and thus
are themselves not active flavours in the proton.
However, at large scales $Q^2 \gg m_c^2$
one can treat the charm quarks as massless partons,
which contribute to the sea.
The charm quark mass $m_c$ is a crucial parameter:
it regulates the ratio of charm and light quarks in the sea
and thus affects predictions
for almost any quark driven process at the LHC.
At LHeC one expects much more precise and kinematically
extended measurements of inclusive charm production compared to HERA.
This will allow to map for the first time the transition from the massive to the massless regime.
Simulations show that one can use 
the data for a $m_c$ determination at a precision
of two per mille.
With very good forward charm tagging one can also test
the hypothesis of an intrinsic charm component in the
proton wave function, which could appear at high $x\simeq 0.2$.
\item
{\bf Beauty:}
Simulation studies show that
one expects at LHeC precise measurements
of inclusive beauty production in DIS.
For large squared momentum transfer $Q^2 \gg m_b^2$
(with $m_b$ being the beauty quark mass)
these measurements can be directly translated
into an effective beauty quark density in the proton.
There is a huge interest in these densities, since many new physics scenarios
involve beauty quarks.
For instance,
in the minimal supersymmetric extension of the standard model
the production of the neutral Higgs boson $A$ is driven by $b\bar{b} \rightarrow A$.
While at HERA the inclusive beauty production results were statistically
limited to about $20\%$ precision, very accurate results
can be expected at the LHeC.
\item
{\bf Top:}
The production of top quarks can be studied at the LHeC for the first time 
in DIS experiments.
The dominant process is single top (or anti-top) production in $W b$ to $t$ fusion.
The unique top physics program that can be performed at
the LHeC includes possibly
the consideration of a quark density for the top, from NC,
a high precision measurement of the top
mass from its decay and cross section.
Top physics at the LHeC is a promising
subject for further study.
\end{itemize}
In summary, while the LHC data can add information to
certain aspects of the quark densities in the proton
using specific reactions (e.g. Drell Yan),
it remains the unique preserve of the LHeC to
completely resolve the quark and antiquark structure of the
proton, for all quark flavours, over the largest
kinematic range ever explored and last but not least with the best theoretical understanding.
%
%

%
%
\subsection{Gluon Distribution}
\label{gluon}

As has been summarised in the CDR, 
there are many fundamental reasons
for the necessity to understand the gluon distribution
and the gluon-parton interactions deeper than hitherto.
Half of the proton's momentum is carried by gluons.
The gluon self-interaction is responsible for the creation of baryonic
mass. In $pp$ scattering at the LHC, the Higgs particle is predominantly
produced by gluon-gluon interactions.
Gluino pair production, as discussed in Section\,\ref{susy},
predominantly proceeds via $gg \rightarrow g \rightarrow \tilde{g} \tilde{g}$
production
and is subject to huge uncertainties at high masses. On the other hand 
the rise of the gluon density
towards low Bjorken $x \lesssim 10^{-5}$ is expected to be tamed
and a new phase of hadronic matter to be discovered, in which
gluons interact non-linearly while $\alpha_s$ is smaller than $1$.

\begin{figure}[htbp]
\centerline{\includegraphics[clip=,width=1.\textwidth]{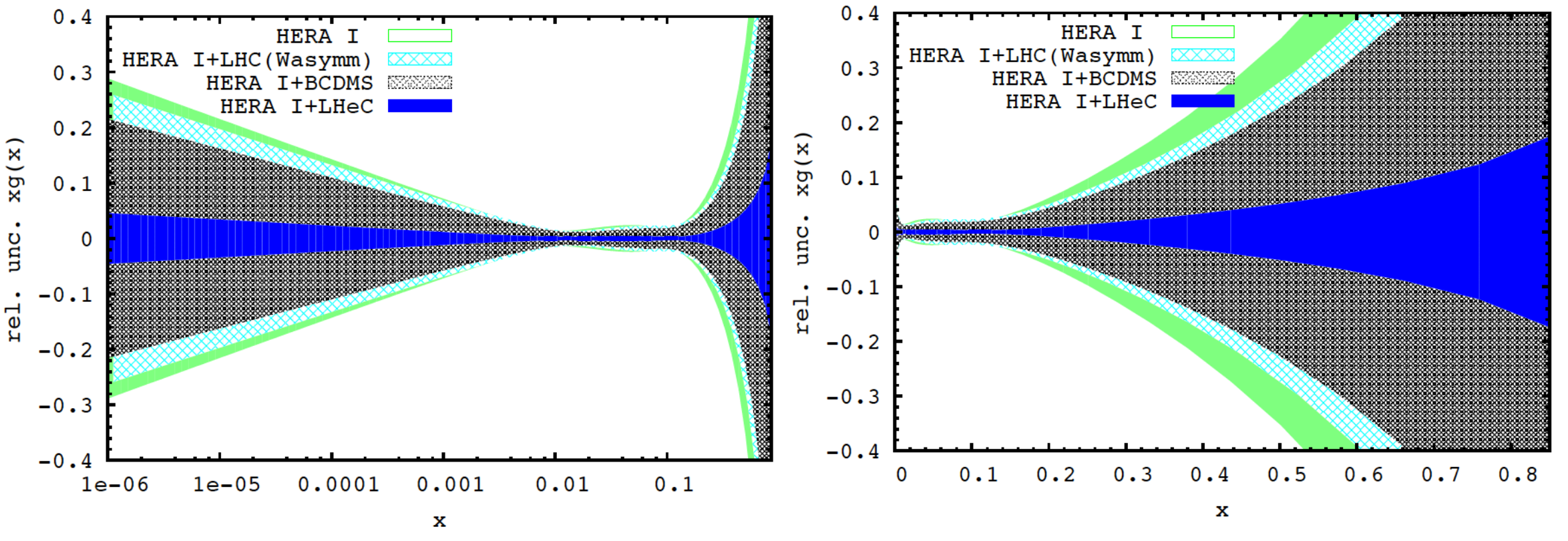}}
\caption{Relative uncertainty of the gluon distribution at
$Q^2 = 1.9$\,GeV$^2$, as resulting from an NLO QCD fit to
HERA (I) alone (green, outer), HERA and BCDMS (crossed),
HERA and LHC (light blue, crossed)
and the LHeC added (blue, dark).
Left: logarithmic $x$ scale, right: linear $x$ scale. 
}
   \label{fig:voiglu}
\end{figure}
\begin{figure}[htbp]
\begin{center}
\includegraphics[clip=,width=0.49\textwidth]{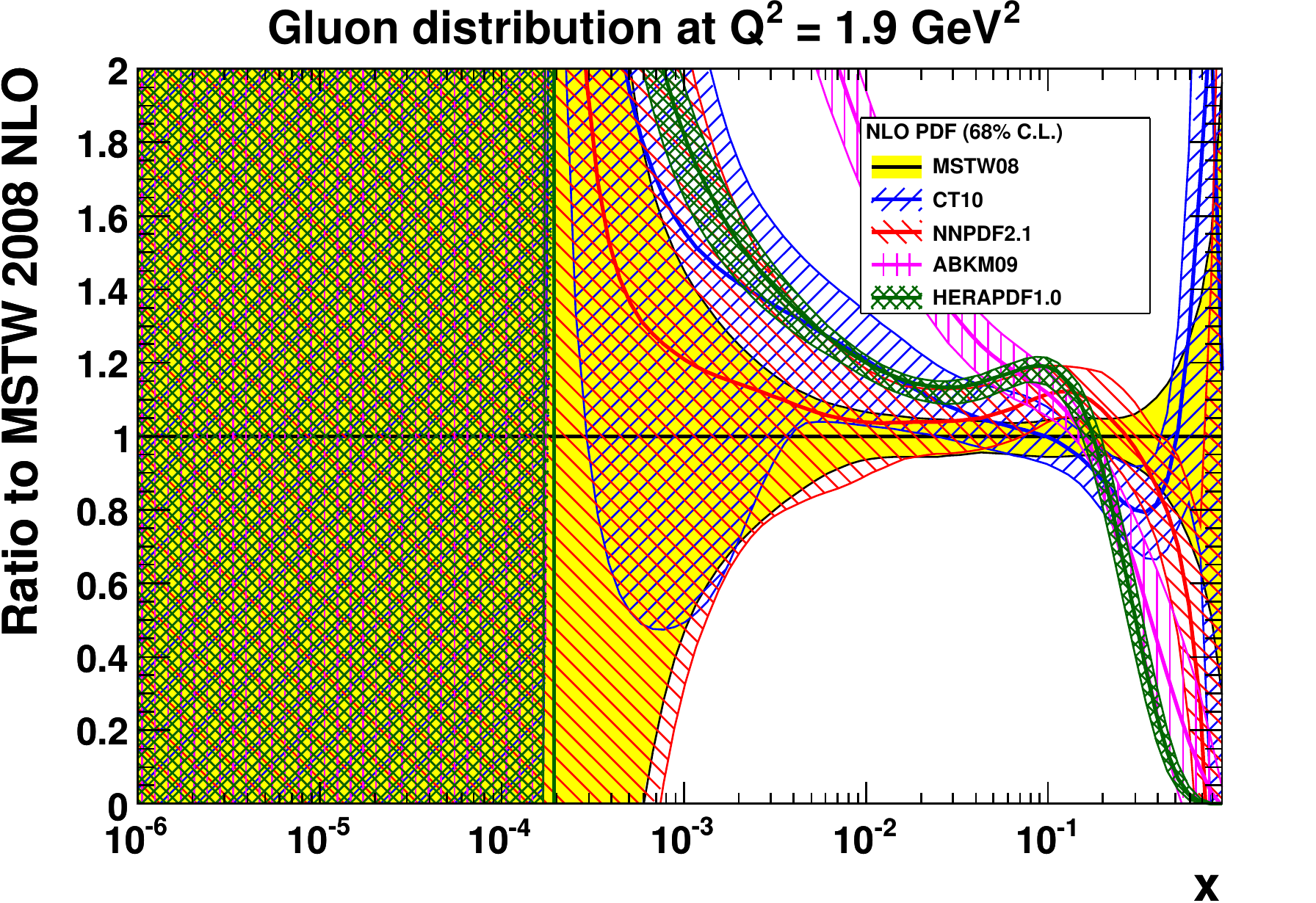}
\includegraphics[clip=,width=0.49\textwidth]{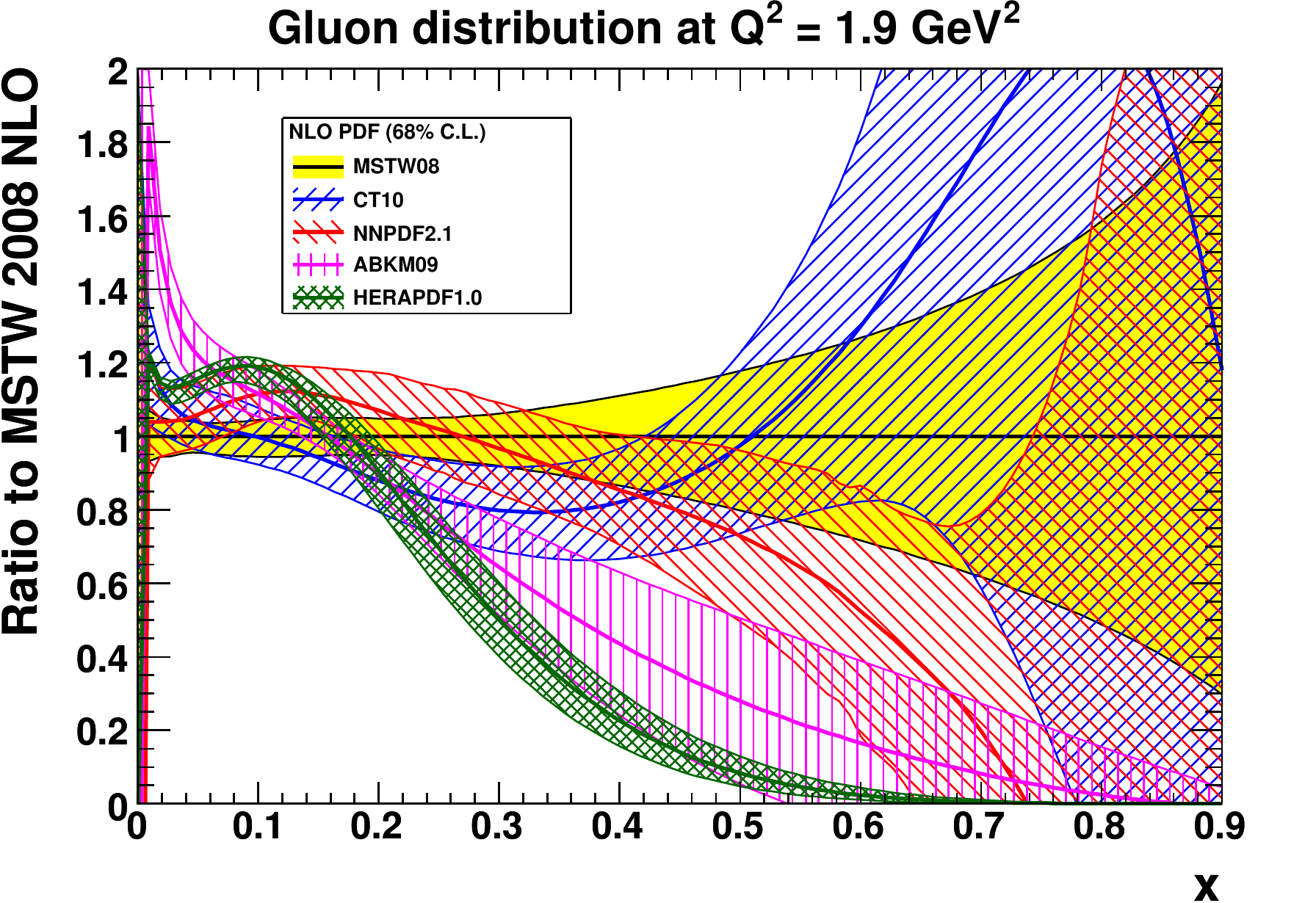}
\end{center}
\caption{Ratios to MSTW08 of
gluon distribution and uncertainty bands, at
$Q^2 = 1.9$\,GeV$^2$, for most of the available recent
PDF determinations, taken from \cite{AbelleiraFernandez:2012cc}.
Left: logarithmic $x$ scale, right: linear~$x$ scale.
}
   \label{fig:graratglu}
\end{figure}

From the simulations as described one derives a typical
uncertainty for the gluon density to be reduced
to about $3,~1,~5$\,\% at $x=5 \cdot 10^{-6},~0.005,~0.5$,
respectively. These values of Bjorken $x$ mark
the low value for saturation to be discovered, the central rapidity
value for Higgs production and the approximate high mass
limit for gluino pair
production. 
As can clearly be seen in Figure\,\ref{fig:voiglu}, 
the potential
of the LHeC for the determination of the gluon density
over $5-6$ orders of magnitude in $x$, simultaneously with
all quark PDFs and $\alpha_s$, is striking. It has to be
compared with the current status
of huge uncertainty on $xg$ at low and high $x$
as is illustrated in Figure\,\ref{fig:graratglu}.
%

%
%
%
\section{PDFs for BSM Searches}
\subsection{Strong Coupling and Grand Unification}
\label{alfas}
Deep inelastic scattering is an ideal process for the determination
of the strong coupling constant, which determines the scaling
violations of the parton distributions. Despite a major effort for
more than $30$ years, there is no precise determination
of $\alpha_s$ available, with a precision competing with QED or weak
couplings, and a number of severe questions remain to be solved as
is discussed in~\cite{AbelleiraFernandez:2012cc} and as has recently 
been summarised in~\cite{job1012}. Questions regard the (in)consistency 
of previous DIS data, the (in)consistency of inclusive DIS and jet based data, the
true uncertainty of the world average on $\alpha_s$ including the
role of various lattice QCD determinations, etc. It is for these
reasons and because of the importance of $\alpha_s$ for the
grand unification of gauge theories, and its importance in a plethora of predictions
of cross sections, such as for Higgs production at the LHC, 
that an experimental determination of the strong coupling
with an order of magnitude improved precision is crucial.
It is also time to challenge the lattice QCD $\alpha_s$ results, which
seem to be most precise but which
exhibit variations which are non-negligible~\cite{job1012}.

Two independent simulations and fit approaches have been undertaken
in order to verify the potential of the LHeC to determine $\alpha_s$,
see the CDR~\cite{AbelleiraFernandez:2012cc} for details. 
Table~\ref{tab:alfa} summarises the main results. It can be seen 
that the total experimental uncertainty on $\alpha_s$ is $0.2$\,\%
from the LHeC and $0.1$\,\% when combined with HERA. This determination
is free of higher twist, hadronic and nuclear corrections relying 
solely on inclusive DIS $ep$ data at high $Q^2$. 
\begin{table}
\begin{center}
\begin{tabular}{|l|c|c|c|}
\hline
case & cut [$Q^2$ (GeV$^2$)] & uncertainty & relative precision (\%)\\
\hline
HERA only & $Q^2>3.5$ & 0.00224 &  1.94  \\
HERA+jets & $Q^2>3.5$  & 0.00099 & 0.82 \\
\hline
LHeC only & $Q^2>3.5$  & 0.00020   & 0.17 \\
LHeC+HERA & $Q^2>3.5$  & 0.00013 & 0.11 \\
LHeC+HERA  & $Q^2>7.0$   & 0.00024   & 0.20 \\
LHeC+HERA  & $Q^2>10.$  & 0.00030 & 0.26  \\
\hline
\end{tabular}
\caption{Results of NLO QCD fits to HERA data (top, without and with jets)
to the simulated LHeC data alone and to their combination,
for details of the fit see ~\cite{AbelleiraFernandez:2012cc}.
The resulting uncertainty includes all the statistical and experimental
systematic error sources taking their correlations into account. 
}
\label{tab:alfa}
\end{center}
\end{table}
There are known further parametric uncertainties in DIS determinations
of $\alpha_s$. These can also safely be expected to be much reduced
by the LHeC, which promises to determine the charm mass, for
example, to a precision of $3$\,MeV, as compared to $30$\,MeV at HERA, corresponding
to an $\alpha_s$ uncertainty of $0.04$\,\%. Matching the
experimental uncertainty requires
that, when the LHeC operates, such analyses must be performed in N$^3$LO
pQCD in order to reduce the scale uncertainty. The ambition to
measure $\alpha_s$ to per mille precision therefore represents
a vision for a renaissance of the physics of deep inelastic scattering
which is a major goal of the whole LHeC enterprise.
Due to the huge range
in $Q^2$ and the high precision of the data, new and decisive tests will
also become available for answering the question of whether the
strong coupling determined with jets and in inclusive DIS
are the same. If confirmed, as is demonstrated in Table\,\ref{tab:alfa}
with the HERA data, a joint inclusive and jet analysis has the 
potential to even further reduce the uncertainty of $\alpha_s$
as simulated here.

\begin{figure}[t]
\unitlength=1cm
\begin{picture}(13,7)
\put(0,0){\includegraphics[clip=,angle=0.,width=.53\textwidth]{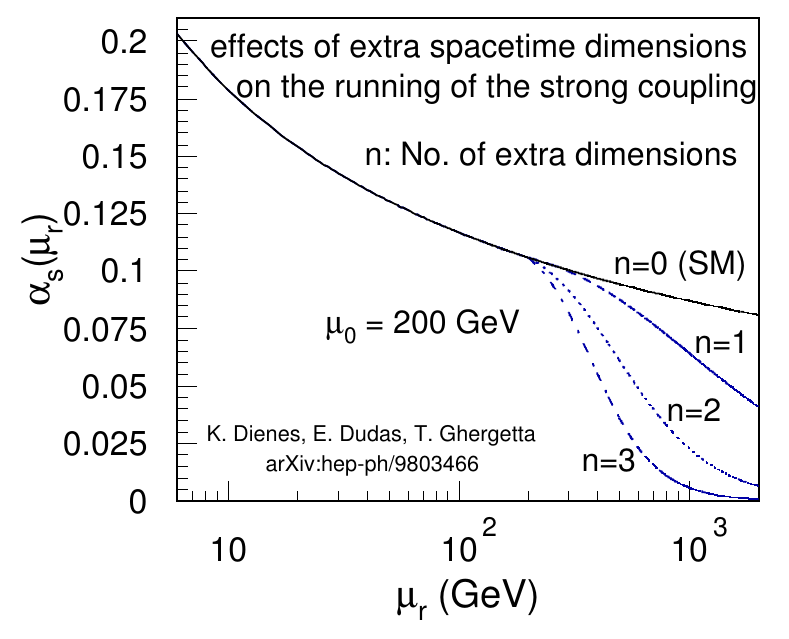}}
\put(9,7.){\includegraphics[clip=,angle=-90.,width=.43\textwidth]{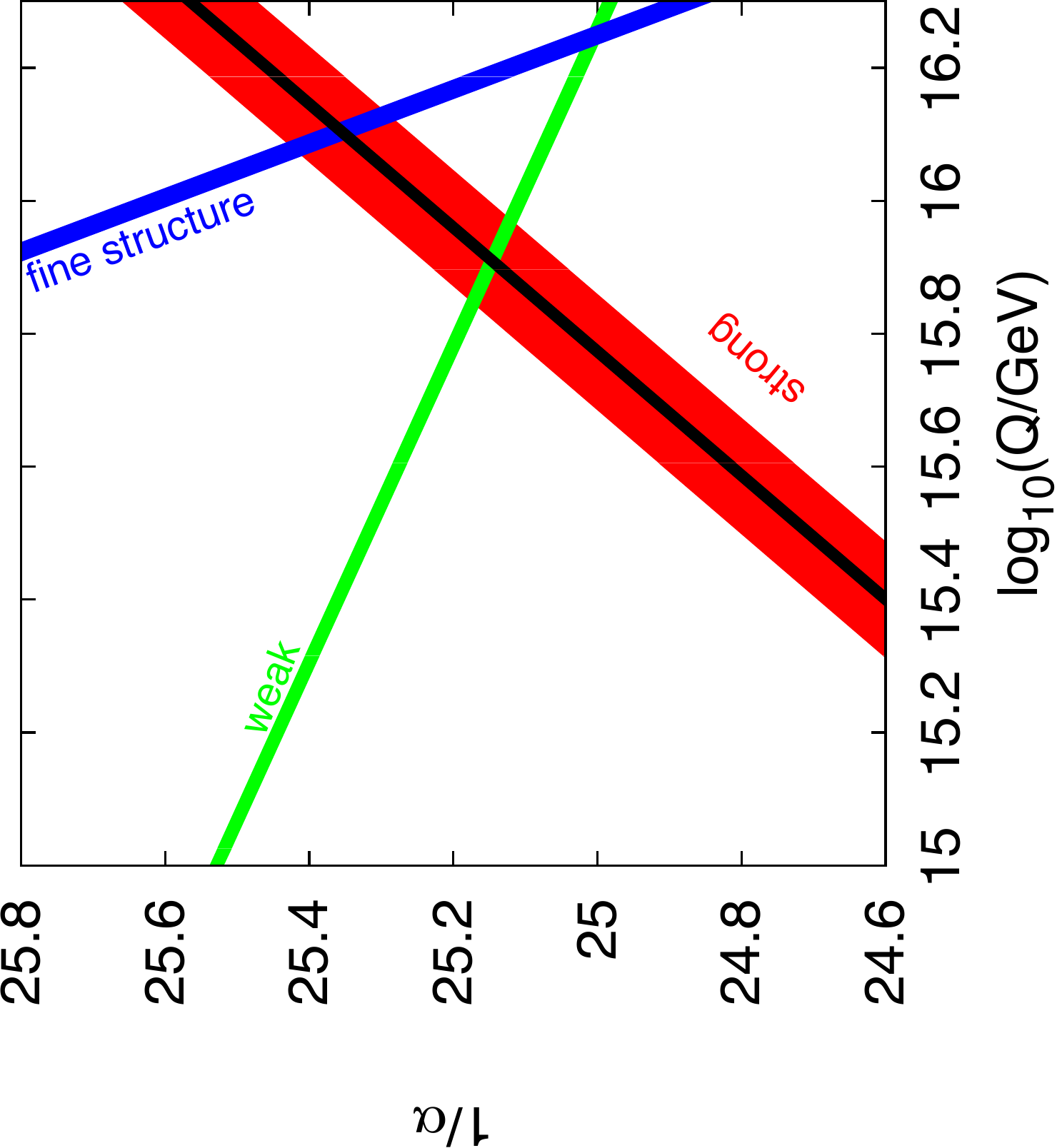}}
\end{picture}
\caption{Left: Running of the strong coupling constant up to high scales
 $\sqrt{Q^2} = \mu$ in the absence (SM) and presence of $n$ extra 
 dimensions~\cite{wobi};
 Right: Extrapolation of the coupling constants (1/$\alpha$) within 
 the MSSM (for the parameter point CMSSM40.2.5~\cite{AbdusSalam:2011fc})
 to the Grand Unified scale as predicted by {\tt SOFTSUSY}~\cite{SOFTSUSY}. The width of the red line is the
 uncertainty of the world average of $\alpha_s$, which
 is dominated by
 the lattice QCD calculation chosen for the PDG average.
 The black band is the LHeC projected experimental 
 uncertainty~\cite{AbelleiraFernandez:2012cc}.
}
\label{alfa}
\end{figure}

Numerous tests of the running of $\alpha_s$ have been performed.
At even higher scales, the law which governs this behaviour would 
be affected, possible strongly as is illustrated in 
Figure\,\ref{alfa} (left), if extra dimensions showed up in the
kinematic range accessible to the LHeC. Besides effects on
$\alpha_s$ one would expect to see changes of the NC cross section,
as in contact interaction patterns, for which the LHeC
provides a range up to about $50$\,TeV, which is discussed in the CDR.

It is well known that grand unified theories (GUTs), having only a single
gauge group,  thus possess a single gauge coupling of that group. 
The Standard Model (SM) gauge couplings are derived, after spontaneous breaking of
the GUT, from the renormalisation group evolution of the gauge couplings, from the
Grand Unified scale to the weak scale where they are measured. Thus there is
a testable GUT prediction: the 
couplings, which are measured at about the weak scale,
should all unify to a common value at
a single, very high energy scale. 
Assuming the SM as the relevant effective field theory, they
do not unify. However, assuming a supersymmetric desert above the weak scale
and using the  
minimal supersymmetric standard model (MSSM), the strong and electroweak
couplings do approximately unify at a common scale of $M_{GUT} \approx 2 \cdot
10^{16}$ GeV.
In the specific calculation used here, the couplings do not quite
match, see Figure\,\ref{alfa} (right).
In realistic GUTs such deviations may occur and are
caused by threshold effects, for example by the prediction of heavy GUT relic
particles that lie just below $M_{GUT}$. 
An accurate inference of this deviation therefore gives important clues into
the structure of such heavy states and therefore, ultimately, the GUT itself. 
It is visible that the present level of uncertainty of the strong
coupling is much larger than that of the weak coupling and the
fine structure constant, while with the LHeC a huge improvement is
expected. 
%

\subsection{Supersymmetry}
\label{susy}
Supersymmetry (SUSY) is a compelling theory providing an 
extension of the Standard Model (SM) at high energies. 
In the SM, the Higgs boson mass suffers from large quantum loop 
corrections, as large as the cut-off scale of the theory, 
and therefore needs a high degree of 
fine-tuning of parameters to be cancelled.
With SUSY, this `naturalness problem' is solved by 
the addition of supersymmetric partner particles to the 
known fermions and 
bosons which 
cancels the largest of these loop effects
and permits the Higgs boson mass to 
lie naturally at the $\sim 100\,$GeV scale. 
Supersymmetric theories present other advantages, 
including the unification of running 
coupling constants at the Planck scale, 
see Sect.\,\ref{alfas}, and renormalisation group 
equations that radiatively generate 
the scalar potential that leads to electroweak symmetry breaking. 
Yet, at the LHC there is so far no sign for SUSY particles.
It is therefore crucial to
increase the beam energy and luminosity
to extend these searches to the limit of phase space.

The possible conservation of $R$-parity, a discrete quantum number which relates 
spin (S), baryon and lepton numbers (B and L), is fundamental in determining the phenomenology of SUSY. 
In the framework of generic $R$-parity conserving supersymmetric extensions 
of the SM, SUSY particles are produced in pairs and the lightest supersymmetric 
particle (LSP) is stable. In a large variety of models the LSP is the 
lightest neutralino, $\tilde{\chi}^{0}_{1}$, one of the SUSY partners of the gauge bosons 
together with its three heavier mass eigenstates ($\tilde{\chi}^{0}_{2,3,4}$)  
and the charginos ($\tilde{\chi}^{\pm}_{1,2}$). The lightest neutralino only 
interacts weakly and provides a  Dark Matter candidate with the appropriate 
relic density to explain the cosmological Dark Matter. 
The possible appearance of $R$-parity-violating couplings, and hence the non-conservation 
of baryon and lepton numbers ($B$ and $L$) in supersymmetric theories, imply an even 
richer phenomenology. Although $R$-parity-violating interactions must be sufficiently 
small, their most dramatic implication is 
the automatic generation of neutrino masses and mixings. The possibility that the results 
of atmospheric and solar neutrino experiments may be explained by neutrino masses and 
mixings originating from $R$-parity-violating interactions has motivated a large number 
of studies and models. 

The discovery (or exclusion) of supersymmetric particles remains a
high priority for the LHC experiments, the LHC being the primary machine to 
search for physics beyond the SM at the TeV scale. The role of the LHeC is to 
complement and possibly resolve the observation of new phenomena.  
At $ep$ colliders, SUSY particles could be produced due to sizeable lepton flavour 
violating terms or, in the framework of $R$-parity conserving models, via 
associated production of selectrons  and first and 
second generation squarks. The latter process 
would present a sizeable cross section only if the sum of selectron and squark masses 
is below or around 1~TeV (thus, for relatively light squarks). 
Current exclusion limits set by the ATLAS and CMS 
experiments on first and second generation squarks are up to 1.5~TeV under the 
assumption that the first two squark families are degenerate. In models where 
this condition is relaxed, windows of discovery relevant for the LHeC 
might still be open at the time of start-up.

Less stringent constraints exist in the context of $R$-parity violating scenarios. 
Processes of interest for $ep$ colliders include leptoquark-like processes  
and associated production of quark-neutralinos via squark-quark-neutralino couplings, 
where squarks can be off-shell and thus beyond direct LHC exclusion limits. While stringent 
constraints exist for associated production of leptons and quarks possibly 
deriving from squark decay, processes of the kind $eu \rightarrow d\tilde{\chi}^{0}_{1}$ 
and subsequent decays of $\tilde{\chi}^{0}_{1}$ to leptons and quarks might be 
difficult to study at the LHC, due to the overwhelming SM multi-jet background, 
and can be successfully searched for at the LHeC.   
The LHeC will also provide indirect handles for the case of supersymmetry.

The dominant SUSY production channels at the LHC are assumed to be 
squark-(anti)squark, 
squark-gluino, and gluino-gluino pair production. All gluon-initiated processes  
suffer from  very large uncertainties due to the extremely limited knowledge
of the gluon density at high $x$.
If gluinos of $2-4$\,TeV mass exist, their discovery and kinematic characterisation 
will depend on the capability to predict their production cross section with 
good precision. 

\begin{figure}[t]
\unitlength=1cm
\begin{picture}(13,7)
\put(3,0){\includegraphics[clip=,angle=0.,width=.6\textwidth]{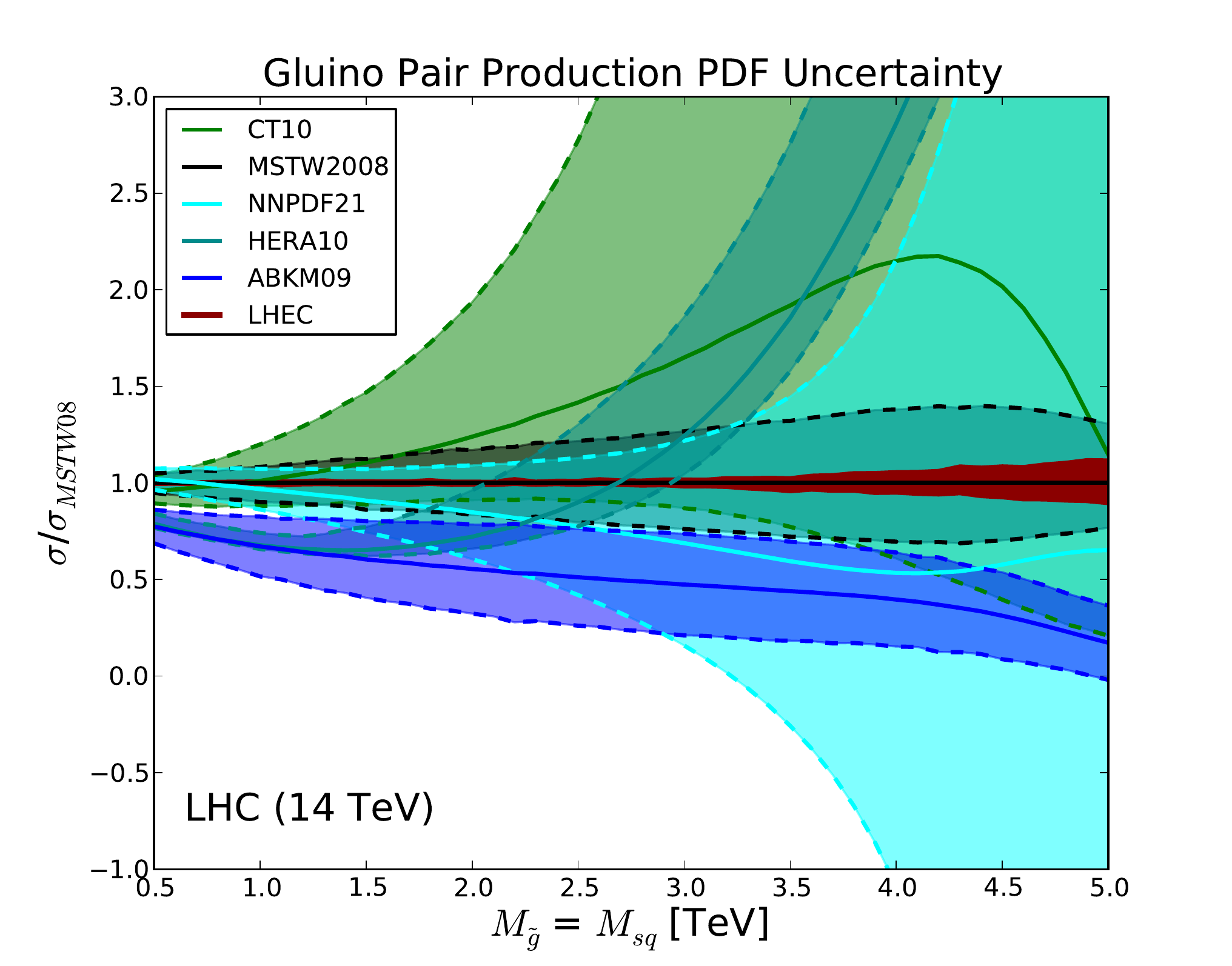}}
\end{picture}
\caption{Calculation of gluino pair production in NLO SUSY-QCD using Prospino~\cite{prospino}
and assuming squark mass degeneracy and equality of squark and gluino masses for illustration.
The error bands are around central values (solid lines) and correspond to
the uncertainty quotations of the various PDF groups. The red band of uncertainty
for the LHeC corresponds to the statistical and systematic errors including their
correlations as treated in the NLO QCD fit described in the CDR.
}
\label{fig:klees}
\end{figure}
Fig.\,\ref{fig:klees} shows the size of PDF 
uncertainties, current and expected, on the gluino-pair production 
cross section calculated at NLO SUSY-QCD expressed 
as the ratio to the MSTW08 predictions. 
The cross section calculation is based on \cite{Beenakker:1996ch} and assumes that first 
and second generation squarks are mass degenerate and equivalent to the gluino mass. 
The renormalisation and factorisation scale is set to the squark/gluino mass. 
Several of the current PDF fits are considered and suffer from very large uncertainties.   
They also differ considerably, by factors, in their central predictions,
which has to do with the smallness of the gluon distribution at large
Bjorken $x$ and the uncertainty of jet physics constraints
at the Tevatron and the LHC, related to scale and calibration
uncertainties and to the size of theory corrections
at high mass Drell-Yan scattering. On the other hand, predictions 
employing the LHeC-derived PDF fits exhibit
much smaller uncertainties, between 5\% and 20\% for gluino 
masses between 500 GeV and 5 TeV, thanks to the expected precision
LHeC measurements of the NC and CC cross sections, and the derived
quark and gluon densities, see above and \cite{AbelleiraFernandez:2012cc}. 
 
Such a greatly reduced level of uncertainty 
is comparable to or below the experimental uncertainties 
expected for squark/gluino searches at the high-luminosity LHC, e.g. through 
enhanced multi-jet production rates (with or without missing transverse momentum) 
compared with SM predictions. It is difficult to predict the
fate of high mass searches for SUSY. However, it is a prime goal of
the GPD LHC experiments in the HL-LHC phase
to explore the phase space up to or close to the
kinematic limit. Fig.\,\ref{fig:klees} makes it clear that
the exploration of the 
region of above a few TeV gluino mass requires a much improved knowledge of PDFs, 
especially for the gluon
for which the LHeC is the most reliable and most precise source, as has been argued above. 
Should deviations from SM predictions be observed, accurate predictions 
for inclusive squark and gluino production cross sections will be crucial to 
understand the nature of the new physics discovered and to determine 
SUSY particle masses and properties~\cite{Dreiner:2010gv}.
In this regard the
LHeC, an ultra-precision QCD instrument, is a
necessary complement for the HL-LHC physics programme,
both for high mass searches as exemplified here and
for making the LHC a precision Higgs factory as
will be illustrated below.

The interest in $R$-parity violating SUSY translates directly 
into the striking potential of the LHeC to determine the
lepto-quark or lepto-gluon quantum numbers should such states
be discovered at the LHC. The $ep$ machine has a clean $s$-channel
single production mode, with variable input beam parameters
while the LHC produces them predominantly in pairs. 
This is discussed in detail in \cite{AbelleiraFernandez:2012cc}, 
along with the reach in contact
interactions, excited leptons, anomalous lepton-quark interactions
and other BSM topics.

\section{Higgs Measurements}
\label{higgs}
\label{sec:higgslhec}
In the Standard Model, the breaking
of the electroweak $SU(2)_L \times U(1)_Y $ symmetry gives mass to the
electroweak gauge bosons via the Brout-Englert-Higgs mechanism,
while the fermions obtain their mass via Yukawa couplings with a scalar Higgs field. 
With the observation of a Higgs-like boson by the 
ATLAS~\cite{:2012gk} and CMS~\cite{:2012gu} 
collaborations with a mass around 126\,GeV, a new research field 
has opened in particle physics. The measurement of the couplings 
of the newly found boson to the known fundamental particles will 
be a crucial test of the SM and a window of opportunity to establish
physics beyond the SM. 

%
%
At the LHeC, a light Higgs boson could be uniquely produced 
and cleanly reconstructed either 
via $HZZ$ coupling in neutral current  DIS or via $HWW$ coupling in 
charged current  DIS. Those vector boson fusion processes have 
sizeable cross sections, O(100)\,fb for $126$\,GeV mass,
and they can be easily distinguished, which is a unique advantage 
in comparison to  the VBF Higgs production in $pp$ scattering.
The observability of the Higgs boson signal at the LHeC was 
investigated in the CDR~\cite{AbelleiraFernandez:2012cc} using 
initially the dominant production and decay mode, i.e.
the CC reaction
 $e^- p \rightarrow H (\rightarrow b \bar{b}) + \nu + X$,
for the nominal $7$~TeV LHC proton beam  
and electron beam energies of $60$ and  
$150\,$GeV.
Simple and robust cuts are identified and found to reject 
effectively e.g. the dominant single-top background,
providing an excellent S/B ratio of about~$1$ at the LHeC,
which may be further refined using sophisticated 
neural network techniques.
At the default electron beam energy of 
$60$\,GeV, for $80$\,\% $e^-$ polarisation and an 
integrated luminosity of 100 fb$^{-1}$, 
the $Hbb$ coupling is estimated to be measurable
with a statistical precision of about 4\,\%, which is
not far from the current theoretical uncertainty.
Typical coupling measurements,
such as $\gamma \gamma$ or $4l$, can be measured with about $10$\,\%
precision with the HL-LHC, while the specific $b\overline{b}$
coupling will be particularly difficult to measure due
to high combinatorial backgrounds in $pp$.

The LHC is said to be inferior to a linear collider
in its coupling measurement prospects. Part of this
statement comes from large uncertainties, which
are related to the imperfect knowledge of the
PDFs and theory parameters. The LHeC, with its 
high precision PDF and QCD programme, will render many
of these uncertainties unimportant. Currently, for example,
an uncertainty of the $H \rightarrow \gamma \gamma$ 
cross section due to PDFs is quoted of nearly $10$\,\%
\cite{Baglio:2012et}, based on the variation of
the cross section predictions from different PDFs.
This will be very much reduced with the LHeC:
an iHix calculation of the NLO Higgs cross section
for MSTW08, NNPDF2.3 and HERAPDF1.5 leads to 
intrinsic uncertainties of 1.7, 1.2 and 2.2\,\%,
respectively, with a maximum deviation of 6.9\,\%.
The full experimental LHeC uncertainty, however, is 0.2\,\%.
The main advantage will be that the precision LHeC
data, possibly combined with HERA, will replace
essentially all previous data sets and thus
lead to a much better agreement between various
PDF determinations, besides the huge reduction
in uncertainty with the LHeC. 

A sizeable uncertainty
is also related to the strong coupling constant,
a difference in $\alpha_s$ of $\pm 0.005$
corresponding to an approximately $10$\,\% cross
section uncertainty, see e.g. \cite{juan}
or \cite{Baglio:2010ae}. Obviously, the large
improvement in the determination of $\alpha_s$
with the LHeC will greatly reduce this uncertainty
too. Essentially with such a high
quality data set as the LHeC can provide, one will simultaneously
determine the coupling and the PDFs, and
control their correlations at a very high level of precision.
In \cite{Denner:2011mq} a systematic evaluation
has been presented of the effect of the
heavy quark masses and of $\alpha_s$ on
the uncertainties of the Higgs branching
fractions in various channels. One finds
sizeable effects, such as $6$\,\% from $M_c$
on the $H \rightarrow c \overline{c}$ branching ratio
or $5.6$\,\% from $\alpha_s$ on $H \rightarrow gg$.
These will certainly be much reduced. It is for
future studies to more systematically analyse
the striking potential of the LHeC to
remove or reduce the QCD uncertainties
on the Higgs cross sections and couplings.
There will also be improvements related to
QCD measurements at the LHC. Their level of
precision, however, especially for the
gluon, $\alpha_s$ and heavy flavour QCD
cannot compete with the DIS data from LHeC.

It has also been observed~\cite{AbelleiraFernandez:2012cc},
 that the LHeC can
specifically explore well the CP structure 
of the $HWW$ coupling by separating it from the $HZZ$ 
coupling and the other signal production mechanisms. Any 
determination of an anomalous $HWW$ vertex will 
thus be free from possible contaminations of these.
A further advantage of the $ep$ collider kinematics stems 
from the ability to disentangle clearly the direction
 of the struck parton and the final state lepton (clear 
definitions of the forward and backward directions). Compared 
to the $pp$ situation, $ep$ lacks 
the complications due to underlying event and pile-up
driven backgrounds.

The few initial studies performed so far will be pursued further
in the light of recent observations from the LHC experiments.
For the projected analyses, this primarily concerns using a full
LHeC detector simulation, and optimising further
the detector design. For the accelerator design it is obvious that
a luminosity in excess of   $10^{33}$~cm$^{-2}$s$^{-1}$
is very desirable, see the discussion on machine
parameters in Section\,\ref{linac}. 
This would open up the possibility of also making precision measurements
of rarer ($\tau$, $Z$, $W$, perhaps photon) 
decay channels, the CP angular distributions for
both the $HWW$ and $HZZ$ couplings, 
and NC initiated production, 
a scenario in which the LHeC collider itself would have a truly remarkable potential to study both the Higgs
boson and mechanism.
Meanwhile, the additional improvements gained in removing a large part of the QCD 
and PDF related uncertainties of the Higgs measurements
in $pp$ will help to make the LHC a precision Higgs facility.

%
%
\section{Heavy Ion Physics}
\label{heavyions}
As discussed in \cite{AbelleiraFernandez:2012cc}, the study of $e$A
collisions at the LHeC will have strong implications on physics of
ultrarelativistic heavy-ion collisions presently studied at RHIC and
at the LHC. This applies both to the initial state which determines
the subsequent behaviour of the dense medium produced in the
collisions, and to the ability of hard probes to characterise such a
medium.  Different physical effects, that are not due to the presence
of a medium, need to be cleanly separated, like nuclear modifications
of parton densities or details of the mechanism of particle
production. Finally, it is also crucial to understand how the medium
affects the probe, for example the mechanism of QCD radiation and
parton energy loss in a medium.

\subsection{Aspects of heavy-ion physics that can be addressed in $e$A at the LHeC}

Currently, nuclear parton distribution functions (nPDFs) suffer from large
 uncertainties  due to the very limited coverage in the kinematics obtained
 by the measurements in low-energy fixed-target experiments. As a result, 
gluon and quark distribution functions are  presently almost unconstrained 
for values of $x$ below $10^{-2}$, which translates into uncertainties in 
the precise characterisation of the medium in heavy-ion collisions through
 hard probes e.g. $J/\Psi$ production \cite{Abelev:2012rv}. 

\begin{figure}[htb]
\begin{center}
\centerline{\includegraphics[clip=,width=.32\textwidth]{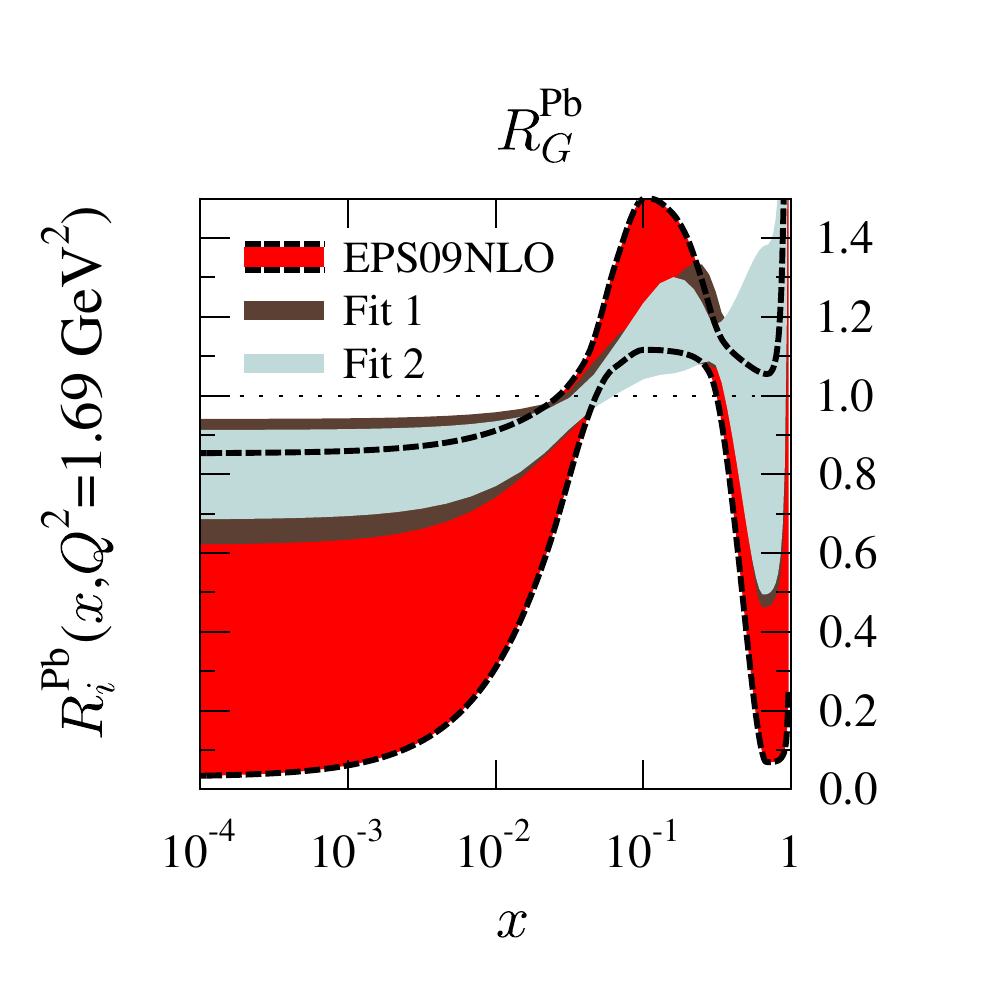}\hskip -0.3cm
\includegraphics[width=0.30\textwidth]{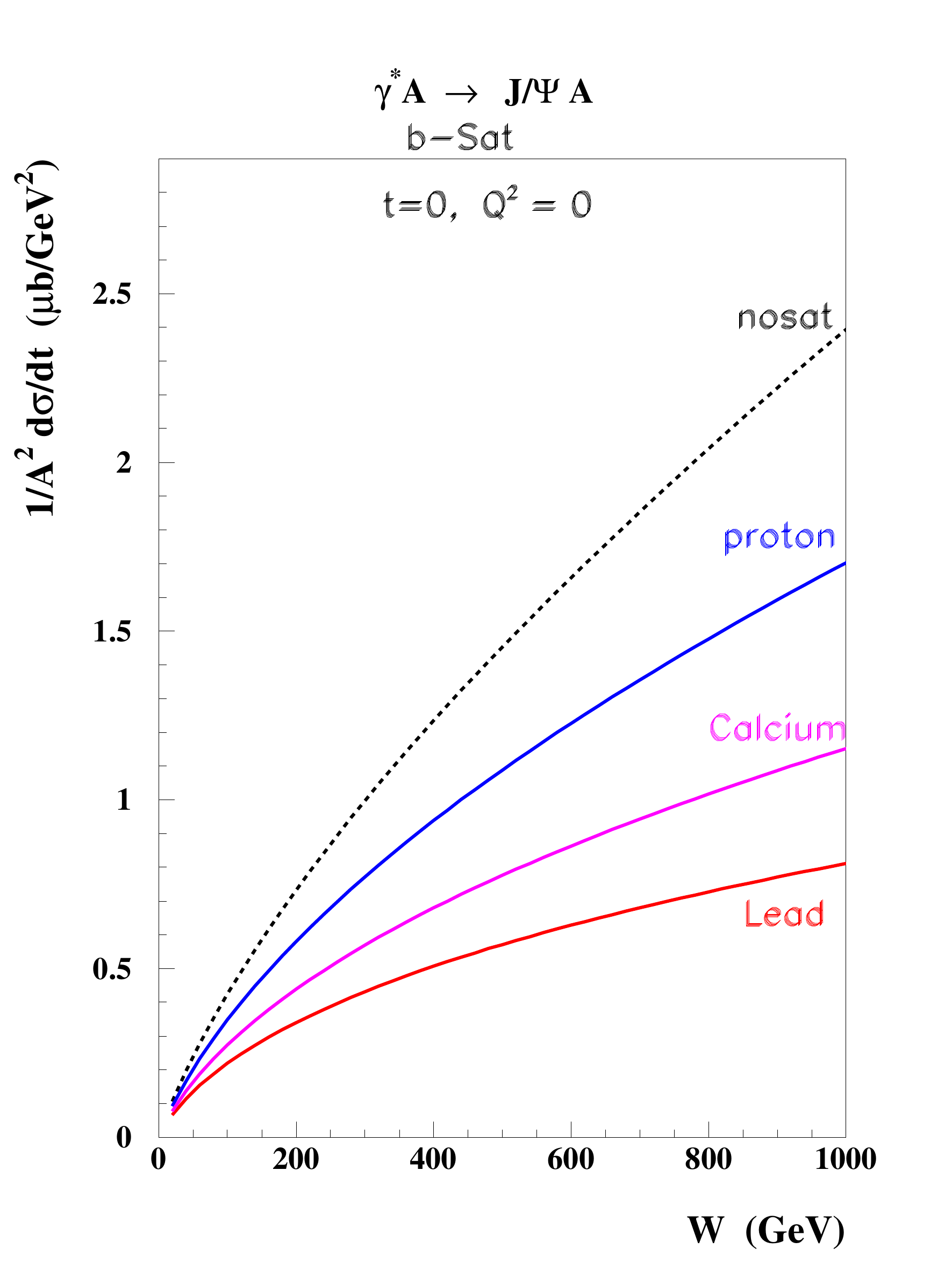}\hskip -0.4cm
\includegraphics[width=0.45\textwidth,angle=0]{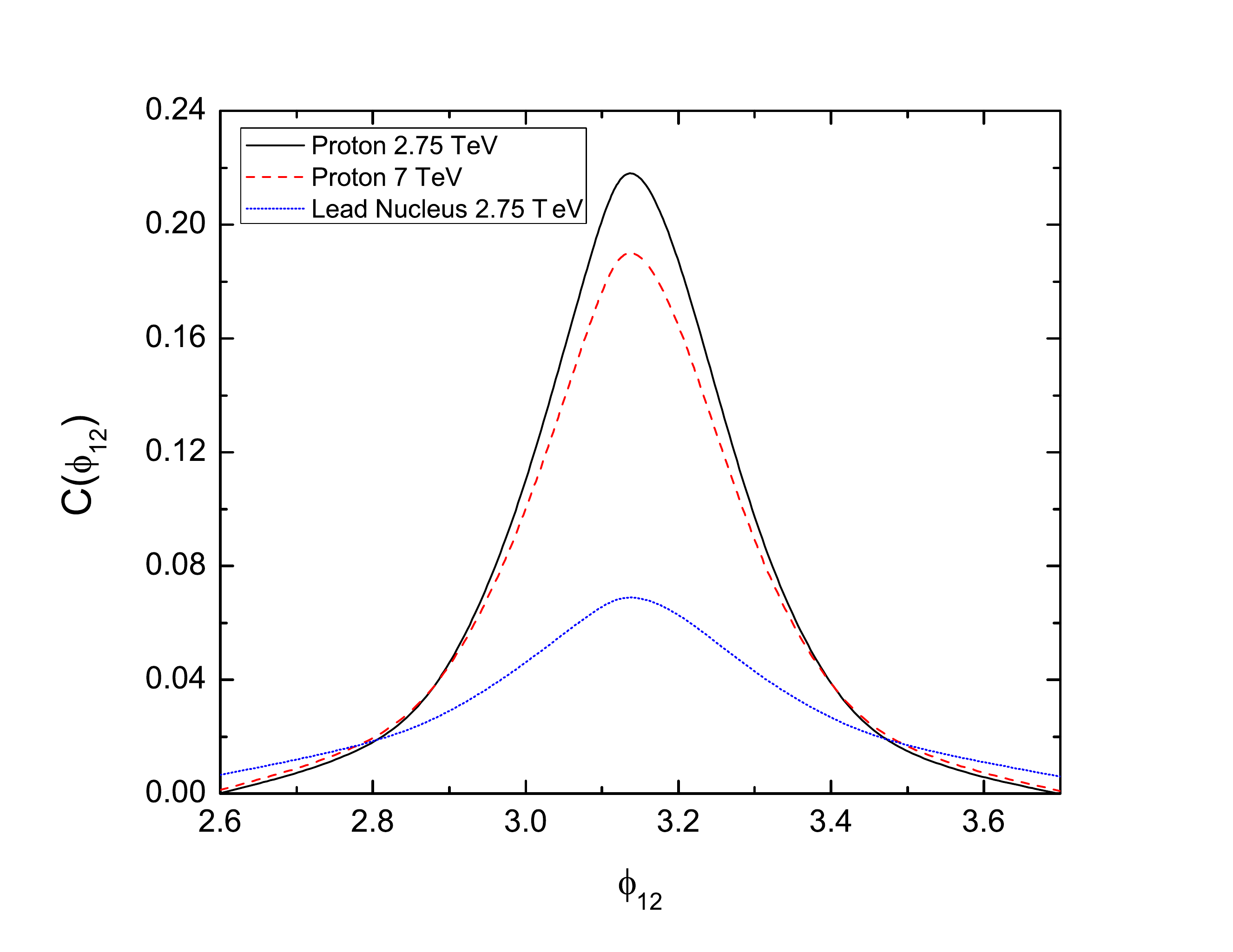}
}
\vskip -0.5cm
\caption{Left: Ratio of gluon density for protons bound in Pb to those in a free proton at $Q^2$ = 1.69 GeV$^2$. The red
 band corresponds to the uncertainty   in the original EPS09 analysis, while the brown band corresponds to the uncertainty
 obtained after including nuclear LHeC pseudodata on the total reduced cross sections (Fit 1).  The light blue band corresponds 
to the uncertainty after including the information on charm  and beauty cross sections (Fit 2).
Centre: Energy ($W$) dependence of the coherent photoproduction of the $J/\Psi$ on a proton and different nuclei in the forward case $t=0$
according to the b-Sat model.  Right: Di-hadron correlation function  for the case of the scattering off the proton 
(red-dashed and black-solid lines) compared to the $e$A case (blue-dotted line). The energy of the electron is assumed 
to be $E_e=50 \  {\rm GeV}$. The observed hadrons are pions.
See \cite{AbelleiraFernandez:2012cc} for details.}
\label{fig:rpdf}
\end{center}
\end{figure}

The LHeC will provide an unprecedented precision for the measurement
of the parton content of nuclei. The kinematic coverage for $e$A
collisions at the LHeC will allow to extend the $x$ and $Q^2$ ranges
by 3 to 4 orders of magnitude down to $x\sim 10^{-5}-10^{-6}$ and up
to $Q^2 \sim 10^6 \; {\rm GeV}^2$ respectively,
see \cite{AbelleiraFernandez:2012cc,cracow1}. Flavour decomposition of
the nuclear structure functions will be performed for the first time,
including the measurement of the previously unknown charm and beauty
components of nPDFs. An example of the constraining power of the LHeC
is illustrated in Fig.~\ref{fig:rpdf} (left) where the nuclear
modification factor is shown for the gluon distribution as a function
of $x$ for a fixed value of $Q^2$.  Clearly, the LHeC offers huge
possibilities for distinguishing between different models of nuclear
shadowing and for constraining the nuclear parton dynamics,
particularly at small values of $x$. Note also that diffraction in
$ep$ and nuclear shadowing are theoretically related through Gribov
relation that can be cleanly tested at the LHeC. Finally, access to
large values of $x > 0.1$ will also be possible at the LHeC, thus
providing additional information about the antishadowing and EMC
effects for nuclear ratios.

Ultrarelativistic heavy-ion collisions at RHIC and the LHC allow to
create and characterise partonic matter under extreme conditions. The
description of the collective behaviour of such matter involves
several stages, which include modelling the initial conditions prior
to isotropisation, the subsequent evolution of the system through
relativistic hydrodynamics and, finally, hadronisation. The initial
conditions for the heavy ion collisions are currently being
parameterised and fitted to the data. They involve large uncertainties
which translate into the uncertainties of the extracted bulk
properties of the medium such as shear viscosity \cite{Shen:2011zc}.
The LHeC offers unique possibilities for pinning down these initial
conditions. The initial state can be determined through the details of
nPDFs and of particle production, both measured precisely in $e$A
collisions.  Furthermore, there are sound theoretical indications that
at low $x$ and for large nuclei, a novel regime of QCD appears that is
characterised by high parton densities. In this regime, the standard
collinear framework that was developed for a dilute parton system must
break down and should be superseded by a more complicated, non-linear
evolution setup that leads to the saturation of parton densities.  The
existence and properties of this partonic dense regime will be tested
at the LHeC through several measurements ranging from those of
structure functions, with subsequent determination of parton
distribution functions in nuclei (see Fig.~\ref{fig:rpdf} (left)) and
inclusive diffraction to more exclusive ones such as elastic vector
meson production (Fig.~\ref{fig:rpdf} (centre)). Thus it will be
possible to locate the onset of the saturation regime as a function of
$x$, mass number and impact parameter of the collision.  Remarkably,
and unlike in lower energy facilities, at the LHeC this phenomenon can
be investigated in the DIS region where the coupling is small and
perturbative techniques are applicable to the non-linear
regime. Furthermore, saturation can be observed both in $ep$ and $e$A,
which gives a unique possibility for disentangling saturation from
other nuclear effects.

Finally, one of the standard tools for characterising the medium
created in heavy-ion collisions is the modification of the yield of
high-energy particles - jet quenching - due to the changes in QCD
radiation and hadronisation induced by the presence of a dense medium.
Such characterisation demands a detailed understanding of these
phenomena that can be achieved by studying particle and jet yields and
correlations in $e$A collisions at the LHeC. In this respect, the
kinematics of the parton whose radiation and hadronisation undergoes
medium modifications is much more precisely constrained in DIS than in
hadronic collisions. As an example, the large kinematic range of LHeC
allows to investigate the dynamics of partons travelling through the
nucleus with energies from moderate to very large.
 Besides, the study of particle angular correlations
 (e.g. back-to-back ones, Fig.~\ref{fig:rpdf} (right)) will offer
 insight both on these changes and on the possible breakdown of
 collinear factorisation and the existence of a dense partonic regime.

\subsection{Prospects in $p$Pb collisions at the LHC compared to $e$Pb at the LHeC}

A first $p$Pb run at the LHC is scheduled for early 2013 (a pilot run
providing ${\cal O}(10^6)$ collisions per experiment happened last
September 13th 2012 and one publication already
appeared \cite{alicehndeta}). Tentative plans exists for additional
runs in the future \cite{jowettcrac}. They offer
information \cite{Salgado:2011wc} about similar aspects relevant for
heavy-ion collisions to those discussed here for $e$A at the LHeC.
The $x-Q^2$ region explored at the LHC with forward instrumentation
both in hadronic and ultraperipheral collisions, and at the LHeC, will
be comparable~\cite{AbelleiraFernandez:2012cc,Salgado:2011wc}

Nevertheless, even assuming that corrections to collinear
factorisation are small - though they are expected to be larger in
$p$A than in $pp$, the nuclear modifications of partonic densities and
of fragmentation and hadronisation, come intrinsically mixed for most
observables in $p$A collisions, see e.g. the discussion
in \cite{Eskola:2012rg}. Therefore, they will be far more difficult to
constrain in $p$Pb at the LHC than in $e$Pb at the LHeC. Besides, the
accuracy - of a few percent - for measurements of cross sections
achievable at the LHeC cannot be matched in a $p$A collider.
Finally, the determination of the kinematic variables relevant for
parton densities, fragmentation functions and other quantities of
interest, is much more direct in DIS than in hadronic
collisions. Thus, the information that can be obtained in $e$Pb
collisions at the LHeC will be substantially more precise, and the
possibilities for discovery of the dense regime of QCD at small $x$
through quantitative studies as those proposed
in \cite{AbelleiraFernandez:2012cc} significantly larger, than in
$p$Pb collisions at the LHC.

%
%
\section{Summary}
The LHeC is a new $ep$ collider of unprecedented kinematic range,
luminosity and precision in deep inelastic scattering.
This leads to the first ever complete measurement
of PDFs, including, for example, the strange density.
It furthermore extends to very high Bjorken $x$ in $ep$ and
to such large $Q^2$ that no nuclear or higher twist
corrections affect the high $x$ PDF determinations.
A much deeper understanding of quark-gluon dynamics
is needed and is in sight, exemplified by the potential of the
LHeC to measure $\alpha_s(M_Z^2)$ to per mille precision.

The LHeC is designed to operate synchronously with the HL-LHC.
Preparations have begun, following the detailed 
design concept report and a corresponding mandate
from CERN, supported by ECFA and NuPECC, to prototype
critical components and to carry out more detailed studies,
such as for the interaction region and high luminosity optics,
in major international collaborations with CERN. As
has been indicated here, there is a potential
for the LHeC to achieve luminosities in excess
of $10^{33}$\,cm$^{-2}$s$^{-1}$. The formation of a
detector collaboration has indeed begun.

The present paper discussed the relation of the LHeC
to the LHC, in its HL phase. As indicated by the 
ATLAS contribution to the ESPP, there are two major
questions for the HL-LHC to investigate, the properties
of the 126\,GeV boson, which likely is the Higgs particle,
and the search for maximum mass particles, of several
TeV in direct production mode. It has been argued here
that the LHeC can assist the transformation
of the  LHC into a precision  Higgs facility,
 with its own coupling
and CP measurements based on clean $WW$ fusion in $ep$,
and with its ultra-precise PDF, heavy quark and
$\alpha_s$ determinations, which will reduce
the theoretical uncertainties of Higgs measurements
in $pp$ to a negligible level. It is similarly 
apparent that the discovery potential of new
particles at high masses at the HL-LHC is severely
limited by the deviations and uncertainties of
the predictions based on currently available
PDF sets, as well as the uncertainties of
input parameters such as the heavy quark masses 
and $\alpha_s$.
In these representative examples, the prominent decay mode
$H \rightarrow b \overline{b}$ has been used for the Higgs, 
while for SUSY, gluino pair production was used.

It has also been discussed which role
DIS and Drell-Yan scattering can play in the 
determination of PDFs, which may be illustrated
by comparing the HERA and Tevatron PDF related results.
Clearly $ep$ provides precision information
on the quark and gluon structure and quark-gluon dynamics,
while the prime task of $pp$ is the extension
of the energy frontier.  The LHeC is the only machine currently proposed
that can provide the data to enable an all-flavour PDF decompsition of 
the proton in an unbiased, assumption-free way.

The paper has also summarised the importance 
of electron-ion scattering for the LHC heavy ion
physics programme, with the determination of nuclear PDFs
in a phase space extended by $4$ orders of magnitude
and diversified with heavy quark nPDFs to be measured
for the first time. The LHeC determines the
initial state of the QGM and leads to a quantitative
understanding of the hadronisation in media, controlled
by the electron DIS kinematics.

The LHeC has its own, fundamental physics programme,
examples being saturation and diffractive physics, and the
resolution of deuteron, neutron and photon structure. 
Its relation to the LHC extends beyond
the currently most prominent examples of Higgs measurements
and SUSY searches, encompassing the top quark, leptoquarks, excited
leptons and electroweak physics. The CDR has demonstrated
that the LHeC may be realised without any major
extra machine delays.
Pairing the unique hadron beams of the LHC with
a new lepton beam would therefore substantially
enrich the physics potential of the LHC facility and make
optimum use of the major investment already made
and envisaged for the LHC.

%
\begin{footnotesize}

\newpage
\section*{LHeC Study Group}
\noindent J.L.Abelleira Fernandez$^{16,23}$, 
C.Adolphsen$^{57}$,
P.Adzic$^{74}$, 
A.N.Akay$^{03}$, 
H.Aksakal$^{39}$, 
J.L.Albacete$^{52}$, 
B.Allanach$^{73}$,
S.Alekhin$^{17,54}$, 
P.Allport$^{24}$, 
V.Andreev$^{34}$
% ...

\bigskip{\it\noindent
$^{01}$ Ankara University, Turkey \\
$^{02}$ SANAEM Ankara, Turkey \\
$^{03}$ TOBB University of Economics and Technology, Ankara, Turkey\\
$^{04}$ LAPP, Annecy, France\\
$^{05}$ University of Antwerp, Belgium\\
$^{06}$ University of Birmingham, UK\\
$^{07}$ INFN Bologna, Italy\\
$^{08}$ IIHE, Universit\'e Libre de Bruxelles, Belgium, supported by the FNRS \\
$^{09}$ University of Buenos Aires, Argentina \\
$^{10}$ INFN Gruppo Collegato di Cosenza and Universita della Calabria, Italy \\
$^{11}$ Massachusetts Institute of Technology, Cambridge, USA\\
$^{12}$ Charkow National University, Ukraine\\
$^{13}$ University of Copenhagen, Denmark \\
$^{14}$ Cockcroft Institute, Daresbury, UK\\
$^{15}$ Rutherford Appleton Laboratory, Didcot, UK \\
$^{16}$ CERN, Geneva, Switzerland\\
$^{17}$ DESY, Hamburg and Zeuthen, Germany\\
$^{18}$ University of Hamburg, Germany\\
$^{19}$ Hampton University, USA \\
$^{20}$ University of California, Irvine, USA \\
$^{21}$ University of Jyv\"askyl\"a, Finland\\
$^{22}$ Kastamonu University, Turkey \\
$^{23}$ EPFL, Lausanne, Switzerland\\
$^{24}$ University of Liverpool, UK\\
$^{25}$ University of California, Los Angeles, USA\\
$^{26}$ Lund University, Sweden\\
$^{27}$ University of Wisconsin-Madison, USA\\
$^{28}$ Universidad Aut\'onoma de Madrid, Spain\\
$^{29}$ University of Mainz, Germany\\
$^{30}$ The University of Manchester, UK \\
$^{31}$ INFN Milano, Italy\\
$^{32}$ University of Milano, Italy \\
$^{33}$ University of Montr\'eal, Canada\\
$^{34}$ LPI Moscow, Russia\\
$^{35}$ Tata Institute, Mumbai, India\\
$^{36}$ Jefferson Lab, Newport News, VA 23606, USA \\
$^{37}$ Brookhaven National Laboratory, New York, USA\\
$^{38}$ Columbia University, New York, USA\\
$^{39}$ Nigde University, Turkey\\
$^{40}$ Budker Institute of Nuclear Physics SB RAS, Novosibirsk, 630090 Russia\\
$^{41}$ Orissa University, India\\
$^{42}$ LAL, Orsay, France\\
$^{43}$ Laboratoire de Physique Th\'eorique, Universit\'e Paris XI, Orsay, France \\
$^{44}$ University of Oxford, UK\\
$^{45}$ CPHT, \'Ecole Polytechnique, CNRS, 91128 Palaiseau, France \\
$^{46}$ UPMC University of Paris 06, Facult\'e de Physique, Paris, France \\
$^{47}$ LPNHE University of Paris 06 and 07, CNRS/IN2P3, 75252 Paris, France \\
$^{48}$ Pennsylvania State University, USA\\
$^{49}$ University of Pisa, Italy\\
$^{50}$ University of Pittsburgh, USA\\
$^{51}$ Charles University, Praha, Czech Republic \\
$^{52}$ IPhT Saclay, France\\
$^{53}$ University of Santiago de Compostela, Spain\\
$^{54}$ Serpukhov Institute, Russia\\
$^{55}$ University of Siegen, Germany \\
$^{56}$ University of Southampton, UK \\
$^{57}$ SLAC National Accelerator Laboratory, Stanford, USA\\
$^{58}$ Tokyo Institute of Technology, Japan\\
$^{59}$ University of Torino and INFN Torino, Italy\\
$^{60}$ NTNU, Trondheim, Norway\\
$^{61}$ KEK, Tsukuba, Japan\\
$^{62}$ Uludag University, Turkey\\
$^{63}$ Uppsala University, Sweden  \\
$^{64}$ TRIUMF, Vancouver, Canada \\
$^{65}$ Paul Scherrer Institute, Villigen, Switzerland\\
$^{66}$ National Center for Nuclear Research (NCBJ), Warsaw, Poland \\
$^{67}$ University of Warsaw, Poland\\
$^{68}$ ETH Zurich, Switzerland\\
$^{69}$ University of Zurich, Switzerland \\
$^{70}$ Max Planck Institute Werner Heisenberg, Munich, Germany \\
$^{71}$ QMW University London, United Kingdom \\
$^{72}$ Laboratori Nazionali di Frascati, INFN, Italy \\
$^{73}$ DAMTP, CMS, University of Cambridge, United Kingdom \\
$^{74}$ University of Belgrade, Serbia \\
$^{75}$ RWTH Aachen University, Germany \\
$^{76}$ Instituto Superior T\'{e}cnico, Universidade T\'{e}cnica de Lisboa, Portugal
}

\bigskip{\it\noindent
$^a$ supported by European Research Council grant HotLHC ERC-2011-StG-279579 and\\
MiCinn of Spain grants FPA2008-01177, FPA2009-06867-E and Consolider-Ingenio 2010 CPAN CSD2007-00042,
Xunta de Galicia grant PGIDIT10PXIB206017PR, and FEDER.\\
$^b$ supported by the U.S. Department of Energy,contract DE--AC02--76SF00515.\\
$^c$ supported by the Academy of Finland, project no. 141555.\\
$^d$ supported by the Sloan Foundation,
DOE OJI grant No. DE - SC0002145 and \\
Polish NCN grant DEC-2011/01/B/ST2/03915.\\
$^e$ supported by the Turkish Atomic Energy Authority (TAEK).\\
$^f$ supported by the P2IO consortium.\\
}

\end{footnotesize}
\end{document}